\documentclass[aps,prb,twocolumn,amsmath,amssymb,nofootinbib,
  superscriptaddress]{revtex4-2}

\usepackage{graphicx}
\usepackage{bm}
\usepackage{amsmath}
\usepackage{amssymb}
\usepackage{physics}
\usepackage[colorlinks=true,linkcolor=blue,urlcolor=blue,citecolor=blue]{hyperref}
\usepackage[usenames,svgnames]{xcolor}
\usepackage{makecell}
\usepackage[normalem]{ulem}
\usepackage{MnSymbol}
\usepackage{changes}

\definechangesauthor[name={RF},color={blue}]{RF}

\newcommand{\Hop}{\hat{H}}

\newcommand{\Top}{\mathcal{T}}
\newcommand{\Topc}{\mathcal{T}_{\mathcal{C}}}
\newcommand{\Himp}{\hat{H}_S}

\newcommand{\Hint}{\hat{H}_{\rm int}}
\newcommand{\Hbath}{\hat{H}_{B}}
\newcommand{\Hhyb}{\hat{H}_{\rm hyb}}
\newcommand{\Heff}{\hat{H}_{\rm eff}}

\newcommand{\HS}{\hat{H}_{\rm S}}
\newcommand{\HB}{\hat{H}_{\rm B}}
\newcommand{\Vop}{\hat{V}}

\newcommand{\Lop}{\mathcal{L}}
\newcommand{\Limp}{\mathcal{L}_S}
\newcommand{\Lint}{\mathcal{L}_{\rm int}}

\newcommand{\aop}{\hat{a}}
\newcommand{\adop}{\hat{a}^{\dagger}}
\newcommand{\bop}{\hat{b}}
\newcommand{\bdop}{\hat{b}^{\dagger}}
\newcommand{\Aop}{\hat{A}}
\newcommand{\Adop}{\hat{A}^{\dagger}}
\newcommand{\boldAop}{\hat{\bm{A}}}
\newcommand{\boldAdop}{\hat{\bm{A}}^{\dagger}}
\newcommand{\bolds}{\bm{s}}
\newcommand{\instrument}{\mathcal{A}}

\newcommand{\AIop}{\hat{A}^I}
\newcommand{\AdIop}{\hat{A}^{\dagger, I}}
\newcommand{\bIop}{\hat{b}^I}
\newcommand{\bdIop}{\hat{b}^{\dagger, I}}

\newcommand{\Xop}{\hat{X}}
\newcommand{\Yop}{\hat{Y}}

\newcommand{\cop}{\hat{c}}
\newcommand{\cdop}{\hat{c}^{\dagger}}
\newcommand{\dop}{\hat{d}}
\newcommand{\ddop}{\hat{d}^{\dagger}}
\newcommand{\bolddop}{\hat{\bm{d}}}
\newcommand{\boldddop}{\hat{\bm{d}}^{\dagger}}

\newcommand{\sgp}{\hat{\sigma}_+}
\newcommand{\sgm}{\hat{\sigma}_-}
\newcommand{\sgx}{\hat{\sigma}_x}
\newcommand{\sgy}{\hat{\sigma}_y}
\newcommand{\sgz}{\hat{\sigma}_z}
\newcommand{\nop}{\hat{n}}

\newcommand{\hc}{{\rm H.c.}}
\newcommand{\rhoop}{\hat{\rho}}
\newcommand{\rhoimp}{\hat{\rho}_S}
\newcommand{\rhobath}{\hat{\rho}_B}

\newcommand{\Zimp}{Z_{S}}
\newcommand{\Zbath}{Z_{B}}

\newcommand{\gK}{\mathcal{K}}
\newcommand{\gI}{\mathcal{I}}

\newcommand{\im}{{\rm i}}
\newcommand{\contour}{\mathcal{C}}

\newcommand{\Trpt}{\widetilde{{\rm Tr}}}

\newcommand{\WI}{{\rm W}^I}
\newcommand{\WII}{{\rm W}^{II}}

\newcommand{\cumulant}[1]{\left\llangle#1\right\rrangle}

\newcommand{\snu}{College of Physics and Electronic Engineering, and Center for Computational Sciences, Sichuan Normal University, Chengdu 610068, China}

\newcommand{\nudt}{College of Science, National University of Defense Technology, Changsha 410073, China}

\newcommand{\jianga}{College of Advanced Interdisciplinary Studies, National University of Defense Technology, Changsha, China}
\newcommand{\jiangb}{Hunan Research Center of the Basic Discipline for Physical States, Changsha, China}

\newcommand{\chena}{Hunan Key Laboratory of Mechanism and technology of Quantum Information, Changsha, China}

\begin{document}

\title{Time-evolving matrix product operators for off-diagonal system-bath coupling}

% \author{Xiansong Xu}
% \affiliation{College of Physics and Electronic Engineering, and Center for Computational Sciences, Sichuan Normal University, Chengdu 610068, China}
% \affiliation{Science and Math Cluster, Singapore University of Technology and Design, 8 Somapah Road, Singapore 487372}

\author{Chu Guo}
\thanks{These authors contribute equally to this work}
% \email{guochu604b@gmail.com}
\affiliation{\nudt}
\affiliation{\chena}

\author{Wei Wu}
\thanks{These authors contribute equally to this work}
% \email{guochu604b@gmail.com}
\affiliation{\nudt}
\affiliation{\chena}

\author{Xiansong Xu}
\affiliation{\snu}

\author{Tian Jiang}
\email{tjiang@nudt.edu.cn}
\affiliation{\nudt}
\affiliation{\jianga}
\affiliation{\jiangb}

\author{Ping-Xing Chen}
\email{pxchen@nudt.edu.cn}
\affiliation{\nudt}
\affiliation{\chena}

\author{Ruofan Chen}
\email{physcrf@sicnu.edu.cn}
\affiliation{\snu}

%\date{\today}


\begin{abstract}
% The time-evolving matrix product operator (TEMPO) method have demonstrated itself as an efficient numerical method to solve the long-time dynamics of bosonic impurity problems. 
% Based on the process tensor framework, we extend the time-evolving matrix product operator (TEMPO) method to solve bosonic quantum impurity problems (QIPs) with off-diagonal system-bath coupling. 
% Our method is a most generic extension of TEMPO, which applies for any QIPs as long as the bath is noninteracting and the system is linearly coupled to the bath.
% It naturally contains all the current developments of TEMPO in more restricted settings.
The time-evolving matrix product operator (TEMPO) method has proven to be a powerful method to study the long-time dynamics of bosonic impurity problems where a small system is linearly coupled to a noninteracting bosonic bath. However, current developments of TEMPO have mostly focused on the case of diagonal system-bath coupling, i.e., $\sum_k \Aop(V_k \bdop_k + \hc)$, with $\Aop$ a Hermitian operator of the system. 
Based on the process tensor framework, we extend TEMPO to the more general case of off-diagonal system-bath coupling in the form $\sum_k (V_k\Aop\bdop_k + \hc)$, where $\Aop$ could be non-Hermitian.
As applications, we study the real-time dynamics of a spin that is coupled to a sub-ohmic bath via the Jaynes-Cummings-type system-bath coupling and compare it against the standard spin-boson model, where we show that the commonly used rotating-wave approximation could be very poor for this bath. We also study the imaginary-time evolution of a bosonic impurity with nonzero on-site interaction that is coupled to a sub-ohmic bath, to illustrate the flexibility of our method.
% Our results show that the commonly used rotating-wave approximation could easily fail in presence of a structural bath.
Our method provides a unified framework to understand different variants of TEMPO, and is a promising building block for an impurity solver in the bosonic dynamical mean field theory for the normal phase with a scalar hybridization function.
% Our method is a natural extension of vanilla TEMPO as its core algorithm can be related to the former without any additional theoretical derivations. 
\end{abstract}

\maketitle

%%%%% introduction

\section{Introduction}
% \textcolor{blue}{Papers mentioned by 2nd referee} \cite{choi2016-exploring,yan2017-equilibration,yan2017-dynamics,bordia2017-probing}.

Understanding the influence of the environment on a quantum system has been a major research topic in quantum theory~\cite{breuer2007-the}, for which two very distinct routes have been taken. In the first route, which is often taken in quantum optics or condensed matter physics, one starts from a microscopic description of the system-environment coupling, and calculates the reduced dynamics of the system afterwards. A prototypical model considered in this route is the spin-boson model~\cite{LeggettZwerger1987}, which describes a two-level spin that is coupled to a continuous noninteracting bosonic bath. 
% and the Anderson impurity model, which describes a localized electron coupled to iteranent electrons~\cite{anderson1961-localized}. 
The spin-boson model is used to study a wide variety of phenomena such as quantum stochastic resonance~\cite{gammaitoni1998-stochasticresonance}, dissipative Landau-Zener transition~\cite{nalbach2009-landau} and quantum phase transition~\cite{kopp2007-universal,winter2009-quantum}.  
In the second route, which is often taken in quantum information, one assumes no (or only partial) prior knowledge of the microscopic model, and tries to algorithmically determine the influence of the environment by measuring the system~\cite{NielsenChuang2010}. 
A seminal progress made in this route is the \textit{process tensor framework}, which provides a complete mathematical foundation for characterizing general non-Markovian quantum dynamics~\cite{CostaShrapnel2016,PollockModi2018a}. The process tensor (PT) contains all the observable information about the system dynamics, which can be systematically and uniquely determined using a procedure similar to quantum process tomography~\cite{WhiteHill2022}. Importantly, PT has a natural matrix product operator (MPO) representation~\cite{PollockModi2018a}, which can be explored for efficient tomography~\cite{Guo2022a,Guo2022c,ZhangGuo2025}.

The TEMPO method~\cite{StrathearnLovett2018} is an intersection of these two routes, which was initially developed for solving bosonic quantum impurity problems (QIPs) in which the system is \textit{diagonally} coupled to the bath via a single Hermitian operator. TEMPO is rooted in the quasi-adiabatic propagator path-integral (QuAPI) method~\cite{makarov1994-path,makri1995-numerical}: QuAPI makes use of the analytical expression of the Feynman-Vernon influence functional (IF)~\cite{FeynmanVernon1963} to trace out the bath and obtain a compact tensor that is responsible for the system dynamics, while TEMPO further represents this tensor with a matrix product state (MPS). 
It was soon realized that the MPS built in TEMPO can be generalized to an MPO that relates to the PT~\cite{JorgensenPollock2019}. Drawing this connection, TEMPO has been extended to study more general QIPs with non-commutative (but still diagonal) couplings to multiple baths~\cite{gribben2022-exact,Chen2024,ZhangShi2025}, and to study optimal quantum control in the presence of non-Markovian quantum noises~\cite{FuxKeeling2021,ButlerEastham2024}.

In this work, we extend TEMPO to the more general case of off-diagonal system-bath couplings where the system is coupled to the bath via a conjugate pair of non-Hermitian operators.
% the most general situation that can allow an analytical expression of the Feynman-Vernon IF. 
We note that some previous works have used off-diagonal coupling to mean that there are several non-commutative system operators and each operator is coupled to its own bath~\cite{Chen2024,ZhangShi2025}, but in this work we use it to mean that there are several non-commutative system operators that are coupled to the \textit{same} bath (and there could also exist several baths).
Similar to TEMPO, our method is only based on the Trotter decomposition of the IF. The crucial difference is that for diagonal coupling, the IF can be interpreted as the partition function of a classical Hamiltonian, which could naturally be represented as an MPS, while the IF for off-diagonal coupling needs to be interpreted as the thermal state of an effective quantum many-body Hamiltonian in our method, which should be represented as an MPO instead.
% but we directly build the IF as an MPO that represents the PT by interpreting it as the thermal state of an effective quantum many-body Hamiltonian. 
Our method naturally reduces to the vanilla TEMPO under the more restricted system-bath couplings. 
We expect our method to have a similar accuracy to TEMPO, but with a moderate computational overhead as MPOs have one more physical index per site compared to MPSs.
We validate our method in two exactly solvable cases: (i) toy models where the baths contain only a single bosonic mode, and (ii) the noninteracting case where the system itself is a noninteracting bosonic mode. As an application, we study the real-time dynamics of a modified spin-boson model with Jaynes-Cummings-type system-bath coupling to a sub-ohmic bath, and compare it against the standard spin-boson model with normal Rabi-type coupling, where we show that the rotating-wave approximation could be very poor for this bath, even if the system-bath coupling is very weak. We also study the imaginary-time evolution of a bosonic impurity with nonzero on-site interaction coupled to a sub-ohmic bath, to illustrate the flexibility of our method and its potential application in the bosonic dynamical mean field theory (BDMFT)~\cite{ByczukVollhardt2008}.

To this end, we briefly discuss the two related works which are also based on a generalization of TEMPO and can deal with the type of off-diagonal system-bath couplings considered in this work. The first one is based on a perturbative treatment of the system path integral, which explicitly depends on the details of the system-bath coupling~\cite{RichterHughes2022}. In comparison our method is blind to these details, similar to original TEMPO and QuAPI, which would be beneficial when considering phenomenological system-bath couplings~\cite{LeggettZwerger1987} or in the BDMFT for studying strongly correlated quantum many-body effects~\cite{choi2016-exploring,yan2017-equilibration,yan2017-dynamics,bordia2017-probing}. 
Another very recent work generalized TEMPO to allow arbitrary number of system operators coupled to the same bath~\cite{Link2026}, which is a more general situation than we have considered in this work. The latter work used a very different strategy from us: they work around the off-diagonal system-bath coupling by going to the eigenbasis representation of each of the system operator (a similar idea was used for non-commutative system operators which are diagonally coupled to different baths~\cite{Chen2024}), and the resultant tensor has a rank that scales linearly with both the number of time evolution steps and the number of non-commutative system operators coupled to the same bath. In our approach, for the conjugate-pair off-diagonal coupling explicitly derived in this work, the rank of the PT is in general determined only by the bath spectral density and time window, and not by the number of coupling operators separately. An extension of our formalism to an arbitrary number $L$ of system operators coupled to the same bath, with a matrix hybridization function $\Delta_{\nu\nu'}$, has been proposed in Sec.~\ref{sec:method} as a conjecture; the corresponding scaling with $L$ remains to be rigorously established. As the generalized IF is the only starting point of our method, it can be straightforwardly applied to the imaginary, Keldysh, or even the L-shaped Kadanoff contours~\cite{kadanoff1962-quantum,AokiWerner2014}, although the last case will not be illustrated in the current work (but this functionality has been supported in our open-source package~\cite{TEMPO}).

% We also note that Ref.~\cite{RichterHughes2022} already proposed a generalization of TEMPO to deal with this scenario. However, their strategy is based on a perturbative treatment of the system path integral, which explicitly depends on the details of the system-bath coupling, while QuAPI and TEMPO are only based on the Trotter decomposition of the IF. The blindness to the details of the system-bath coupling would be beneficial when considering phenomenological system-bath couplings~\cite{LeggettZwerger1987} or in the bosonic dynamical mean field theory (BDMFT)~\cite{ByczukVollhardt2008}. 
% A recent work generalized TEMPO to allow arbitrary number of impuriy operators coupled to the same bath, which is a more general situation than we have considered in this work~\cite{Link2026}. The latter approach has used a very different derivation from us: they

The paper is organized as follows. 
In Sec.~\ref{sec:pt}, we briefly review the process tensor framework from a computational perspective.
In Sec.~\ref{sec:method}, we present our extended TEMPO method and discuss its connection to the previous developments of TEMPO.
% In Sec.~\ref{sec:processtensor}, we show that the discretized IF is exactly the PT, and naturally reduced to TEMPO and its existing extensions for more restricted system-bath couplings.
 % is naturally review the process tensor framework and explain its intmate connecton with the Feynman-Vernon IF for the spin-model model. 
% In Sec.~\ref{sec:method}, we explain the intmate connecton between the process tensor and the Feynman-Vernon IF for general spin-boson model.
In Sec.~\ref{sec:results} we present our numerical results. We summarize in Sec.~\ref{sec:summary}.

\section{Brief review of the process tensor framework}\label{sec:pt}
\begin{figure}
  \includegraphics[]{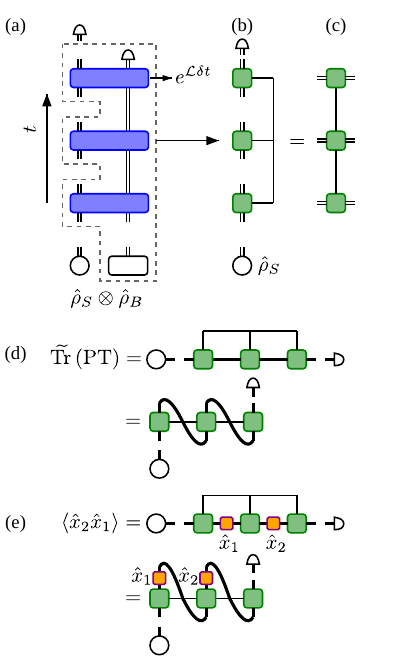} 
  \caption{(a) Discretized time evolution of a system $S$ coupled to an environment $E$ under an evolutionary operator $e^{\Lop \delta t}$, starting from a separable initial state $\rhoop_0=\rhoop_S\otimes \rhoop_E$. We have used double leg for the vertical (physical) indices as the Liouville (or more generally Lindblad) equation of motion is used, which evolves the density matrix instead of the pure state. The empty arcuate hat means the inner trace of the double leg.
  The regime within the dashed box defines the process tensor, which is the tensor network shown in (b), and can be equivalently viewed as an MPO in (c). (d) Trace of the PT, which is $1$ if $\rhoop_0$ is properly normalized. (e) Calculating multi-time correlations based on the PT. In (d,e), we have used a thicker leg to represent the double leg for brevity. 
  % The contraction with the system initial state and the final trace over the system is only taken when calculating observables, such that the MPO form of the PT can be 
    }
    \label{fig:pt}
\end{figure}

% The spin-boson model is a special instance of the broader class of quantum impurity problems (QIPs), which describes a small impurity (the system) that is coupled to a noninteracting bath of free oscillators.
A basic requirement when studying QIPs is to calculate the multi-time correlation functions of the system. The process tensor framework provides a theoretical foundation for such calculations. We consider a discretized time evolution of a quantum system $S$ coupled to some (possibly unknown) environment $E$ (in this work, bath is used for free oscillators while environment can be any general quantum degrees of freedom) under an evolutionary operator $e^{\Lop \delta t}$ with step size $\delta t$ ($\Lop$ can be a Liouville operator or more generally a Lindblad operator), intertwined by some quantum operations at each time step which are denoted as $\instrument_k$. 
% We further assume that the initial state $\rhoop_0$ is separable, i.e, $\rhoop_0 = \rhoop_{S}\otimes \rhoop_E$. 
Then the reduced density matrix of the system after $k$ steps of evolution can be written as
\begin{align}\label{eq:ptdef}
\rhoop_S(k\delta t) = \Tr_E\left( \instrument_k \cdots e^{\Lop \delta t} \instrument_2 e^{\Lop \delta t} \instrument_1 \rhoop_0 \right).
\end{align}
Here $\rhoop_0$ denotes the $SE$ initial state, which can be any entangled state, but in the context of this work we only consider the case of a separable initial state. $\Tr_E$ means the partial trace over $E$. Eq.(\ref{eq:ptdef}) defines a $k$-step PT, which is a linear mapping from $k$ quantum operations $\{\instrument_1, \cdots, \instrument_k\}$ into $\rhoop_S(k\delta t)$~\cite{CostaShrapnel2016,PollockModi2018a}. PT is schematically illustrated in Fig.~\ref{fig:pt}(a). Importantly, it can be naturally represented as an MPO, as shown in Fig.~\ref{fig:pt}(b,c), and the bond dimension of which can be interpreted as the minimal memory size required to generate the observed system dynamics. 
If both $\rhoop_0$ and $e^{\Lop \delta t}$ are known, then PT is exactly known from Fig.~\ref{fig:pt}(a,b,c). Importantly, even if one has completely no prior knowledge of them, PT can still be obtained using process tensor tomography as long as one has access to measure any system observables~\cite{WhiteModi2020,WhiteHill2022}.
Once PT is obtained, any multi-time correlations of the system can be calculated by properly taking the trace operation on it, as shown in Fig.~\ref{fig:pt}(d,e). Here we have used $\Trpt$ to denote the trace of PT, as it differs from the standard trace operation on MPOs~\cite{Schollwock2011}.

\section{Method}\label{sec:method}

\subsection{Description of bosonic QIPs}

In this work we focus on bosonic QIPs. The total Hamiltonian of a bosonic QIP can be written as
\begin{align}\label{eq:H}
\Hop = \Himp + \Hint,
\end{align}
where $\Himp$ is the system Hamiltonian, $\Hint = \Hhyb + \Hbath$ contains the bath Hamiltonian $\Hbath$ and the coupling $\Hhyb$ between system and bath. We will use $d$ to denote the local Hilbert space size of the impurity (e.g., $d=2$ for a two-level spin).
Two assumptions are generally made for QIPs: (i) the bath is noninteracting, i.e., $\Hbath = \sum_{l,k} \omega_{l,k} \bdop_{l,k}\bop_{l,k}$ where $\bdop_{l,k}$ and $\bop_{l,k}$ are creation and annihilation operators for the $k$th mode of the $l$th bath (the system could be simultaneously coupled to multiple baths in general), and (ii) the system is linearly coupled to the bath, i.e., $\Hhyb$ is linear in terms of $\bdop_{l,k}$ and $\bop_{l,k}$.

The original TEMPO has been restricted to diagonal and commutative system-bath coupling, i.e., $\Hhyb = \sum_{l, k} \Aop_l (V_{l, k} \bdop_{l, k} + \hc)$, in which there is only a single Hermitian operator $\Aop_l$ of the system that is coupled to the $l$th bath (diagonal), and $\Aop_l$ commutes with each other (commutative)~\cite{StrathearnLovett2018}. Later TEMPO was extended to the case of diagonal but non-commutative system-bath coupling in which $\Aop_l$ does not commute with each other~\cite{gribben2022-exact,Chen2024,ZhangShi2025}. In this work we consider the more general case of off-diagonal system-bath coupling with
\begin{align}\label{eq:Hhyb}
\Hhyb = \sum_{l, k}  (V_{l, k} \Aop_l \bdop_{l, k} + \hc),
\end{align}
where $\Aop_l$ is not Hermitian and does not commute with each other in general. 
We note that off-diagonal coupling cannot be reduced to diagonal and non-commutative coupling in which the system is coupled to $M$ baths via $M$ non-commutative Hermitian operators, each diagonally coupled to one bath. For example, one could use $\Aop_l =\Xop_l -\im \Yop_l$ and $\Adop_l = \Xop_l +\im \Yop_l$ to convert the above $\Hhyb$ into
\begin{align}\label{eq:Hhyb2}
\Hhyb = \sum_{l, k}  [\Xop_l (V_{l, k}\bdop_{l, k} + \hc) + \Yop(-\im V_{l, k}\bdop_{l, k} + \hc)],
\end{align}
where $\Xop_l$ and $\Yop_l$ are Hermitian operators by definition. But in Eq.(\ref{eq:Hhyb2}) there will be two non-commutative system Hermitian operators that are coupled to each bath, which is not diagonal from our definition.

% \subsection{Process tensor for QIPs}

\begin{figure}
  \includegraphics[]{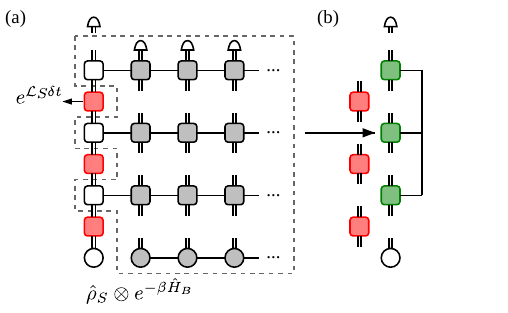} 
  \caption{(a) Schematic illustration of the real-time evolution of the QIP from a separable system-bath initial state as a $1+1$D graph, where the bottom row means the initial state, the rows above mean the propagator $e^{\Lint \delta t}$, the red square means the bare system propagator $e^{\Limp \delta t}$. The regime within the dashed box contains all the influence of the bath on the system dynamics, corresponding to that in Fig.~\ref{fig:pt}(a), which is naturally an MPO as shown in (b).
    }
    \label{fig:qiptn}
\end{figure}

% one does not allow several $\Aop_i$ of the impurity to couple to the same bath. 

Now we show the crucial role that PT can play in solving QIPs.
We denote the Liouville operator as $\Lop$, which acts on the total density operator $\rhoop$ as $\Lop \rhoop = -\im (\Hop \rhoop - \rhoop\Hop)$, and we consider a separable system-bath initial state:
\begin{align}\label{eq:rho0}
\rhoop_0 = \rhoimp \otimes \rhobath,
\end{align}
where $\rhoimp$ is an arbitrary initial state of the system, and $\rhobath = e^{-\beta \Hbath}$ denotes the (unnormalized) thermal state of the bath with $
\beta$ the inverse temperature (when there are multiple baths, their temperatures are allowed to be different).
% Similar to $\Lop$, we define the Liouvillian operators $\Limp$ and $\Lint$ corresponding to $\Himp$ and $\Hint$. 
Performing a first-order Trotter decomposition of the evolutionary operator: $e^{\Lop\delta t}\approx e^{\Lint\delta t}e^{\Limp\delta t}$, where the Liouville operators $\Limp$ and $\Lint$ are defined similar to $\Lop$ and correspond to $\Himp$ and $\Hint$ respectively, then the discretized time evolution of the problem can be understood using the $1+1$D graph as shown in Fig.~\ref{fig:qiptn}(a). 
% The bottom row corresponds to the initial state $\rhoop_0$, and each row above corresponds to $e^{\Lint \delta t}$. 
The continuous bath in the original problem can either be directly discretized into a finite number of modes~\cite{InesWolf2015}, in which bath discretization error would occur, or using the more sophisticated pseudo-mode mapping which could be made exact in principle~\cite{TamascelliPlenio2018,ParkLin2024}.
In the tensor network (TN) language, one could represent the initial state as an MPS, the evolutionary operator $e^{\Lint \delta t}$ as an MPO, then Fig.~\ref{fig:qiptn}(a) would represent a 2D TN, and the problem could be solved by contracting it. In fact, this is the central idea used in the t-MPS method to solve QIPs, which contracts the 2D TN using the boundary MPS method~\cite{VerstraeteCirac2004b} from bottom up and stores the system-bath state at each time step as an MPS~\cite{ChinPlenio2010,PriorPlenio2010,InesBanuls2015,GuoPoletti2018,ChenPoletti2020,HuangGuo2020}.

The process tensor framework inspires an alternative strategy to contract the 2D TN in Fig.~\ref{fig:qiptn}(a): since one is usually not interested in the bath, one could calculate the PT which contains all the information of the system only, instead of calculating the system-bath state. To obtain the PT, one could contract the TN from right to left until only the system degrees of freedom are left, resulting in a single MPO as shown in Fig.~\ref{fig:qiptn}(b). The MPO contains all the influence of $\Hint$ on the system dynamics. 
The influence of $\Himp$ can be simply absorbed into this MPO by applying a local operation $e^{\Limp\delta t}$ on each physical index of it, which will result in the PT of the system. One could also absorb $e^{\Limp\delta t}$ into $e^{\Lint \delta t}$ to obtain an MPO for $e^{\Lop\delta t}$ at the very beginning. However, in practice it could be advantageous to leave $e^{\Limp\delta t}$ uncontracted, as shown in Fig.~\ref{fig:qiptn}(b), since then it will be very flexible to consider any $\Himp$ without recalculating the MPO for $\Hint$. 
Similarly, we will leave the initial state $\rhoimp$ and the final output leg of the system uncontracted until one calculates the system observables, which could allow one to preserve the standard form of MPO and to flexibly choose any initial state $\rhoimp$ and apply any operation at the final step, as shown in Fig.~\ref{fig:qiptn}(b). 
We note that the above strategy is completely general for any quantum many-body problems beyond the QIPs.
It was first proposed in Ref.~\cite{GuoPoletti2020} and illustrated on a toy model with a few spins, and was later applied to solve the real-time dynamics of large-scale 1D quantum many-body problems~\cite{YeChan2021,CygorekGauger2022}. As in Ref.~\cite{CygorekGauger2022}, this strategy will be referred to as the automated compression of environments (ACE) method.

A common pitfall of both the t-MPS and ACE methods is that the bath degrees of freedom have to be explicitly dealt with, which would unavoidably suffer from the bath discretization error and the local bosonic Hilbert space truncation error, especially in situations where the bath modes could get highly excited (here we note the local basis optimization technique to leverage the latter issue~\cite{GuoVojta2012}). In contrast, the method presented below avoids these errors by working directly with the analytic hybridization function, but it introduces its own controllable error sources (contour discretization, exponential representation of the hybridization function, XTRG construction, and MPO compression), which are systematically analyzed in Sec.~\ref{sec:results}.

\subsection{Feynman-Vernon IF for off-diagonal system-bath coupling}

For QIPs, a crucial observation is that the PT does not have to be obtained by discretizing the bath and then using the ACE method, since
the Feynman-Vernon IF already provides an analytic expression for all the influence of $\Hint$ in the continuous-time limit. To illustrate this, we first consider the situation of off-diagonal system-bath coupling to a single bath, i.e.,
\begin{align}\label{eq:jccoupling}
\Hhyb = \sum_k(V_{k} \Aop \bdop_{k} + \hc).
\end{align}
The generalization to the case of multiple baths will be discussed later.
In this situation, the system path integral $\Zimp(t)$ for the real-time evolution, defined as $\Zimp(t) = \Tr\left(e^{\Lop t}\rhoop_0 \right)/\Zbath$ ($\Zbath=\Tr(\rhobath)$ is the partition function of the bath), could be formally written as (See Appendix.~\ref{app:derivation} for the detailed derivation of this equation):
% \begin{align}\label{eq:Zimp}
% \Zimp(t)  =\Tr_S\left[\Top \langle e^{\int_{0}^t \Lint(\tau)  \dd \tau }\rangle_B e^{\int_{0}^t \Limp(\tau) \dd \tau} \rhoimp \right] ,
% \end{align}
\begin{align}\label{eq:Zimp}
%\Zimp(t)  =\Trpt\left[\Topc \gI\boldAdop,\boldAop]\gK[\Himp] \right] ,
\Zimp(t)  =\Trpt\left[\gI[\boldAdop,\boldAop]\gK[\Himp] \right] ,
\end{align}
where $\boldAdop=\{\Adop(t)\}$ and $\boldAop=\{\Aop(t)\}$ denote the operator paths on the Keldysh contour $\contour=0\rightarrow t\rightarrow 0$~\cite{Keldysh1965,LifshitzPitaevskii1981,AokiWerner2014,Chen2025}.
%where $\contour$ denotes the Keldysh contour $0\rightarrow t\rightarrow 0$~\cite{Keldysh1965,LifshitzPitaevskii1981,AokiWerner2014,Chen2025}, $\boldAdop=\{\Adop(t)\}$ and $\boldAop=\{\Aop(t)\}$ denote the operator paths on $\contour$, $\Topc$ denotes the contour-ordering operator which orders the terms with preceding time on $\contour$ to the right. 
% where $\langle \bullet \rangle_B = \Tr_B (\bullet\rhobath)$. The time dependence in $\Limp(\tau)$ and $\Lint(\tau)$ is to explicitly stress that they only act at time $\tau$. $\Top$ is the time-ordering operator will arrange the operators with smaller times to the right.
% After discretization, the first term inside the square braket of Eq.(\ref{eq:Zimp}) would correspond to the influence of $\Hint$, as indicated by the dashed box in Fig.~\ref{fig:qiptn}(a), while the second term correspond to the influence of $\Himp$, as indicated by the red squares in Fig.~\ref{fig:qiptn}(a). 
The first term in the square bracket of Eq.(\ref{eq:Zimp}) is the contour-ordered Feynman-Vernon IF that encodes the influence of $\Hint$:
\begin{align}\label{eq:IF}
\gI[\boldAdop, \boldAop] = \Topc \, e^{-\int_{\contour}\dd \tau \int_{\contour} \dd \tau' \Adop(\tau) \Delta(\tau, \tau') \Aop(\tau')},
\end{align}
where the time dependencies in $\Adop(\tau)$ and $\Aop(\tau)$ are only to stress that they act at time $\tau$, and $\Topc$ denotes the contour-ordering operator that orders operators according to their position on the contour $\mathcal{C}$.
Formally, the operator-valued IF may be interpreted as acting on a ``history Hilbert space''~\cite{gribben2022-exact} in which each point on the contour carries its own copy of the system Hilbert space, so operators at different contour times commute. In this picture, the contour-ordering operator $\Topc$ is redundant and can be omitted. The only ambiguity happens at the point $\tau=\tau'$, which is not an issue in the continuous limit as its measure is zero, but could matter for finite discretization of the IF and is resolved in Appendix.~\ref{app:quapi}.
$\Delta(\tau, \tau')$ is the hybridization function that can be calculated as
\begin{align}\label{eq:hybridization}
\Delta(\tau, \tau') = \im \int \dd\omega J(\omega) D_{\omega}(\tau, \tau'),
\end{align}
with $J(\omega) = \sum_k \abs{V_k}^2\delta(\omega-\omega_k)$ the bath spectral function (BSF), and $D_{\omega}(\tau, \tau')=-\im\expval{T_{\mathcal{C}}\bop_{\omega}(\tau)\bdop_{\omega}(\tau')}_B$ the free bath contour-ordered Green’s function.
The second term in the square bracket of Eq.(\ref{eq:Zimp}) denotes the contributions of $\Himp$ and $\rhoimp$, which can be explicitly written as
\begin{align}\label{eq:gK}
\gK[\Himp] = \Topc \, e^{-\im \int_{\contour} \dd\tau \Himp(\tau)} \rhoimp ,
\end{align}
where we have used the time dependence in $\Himp(\tau)$ again to stress that it acts at time $\tau$. The $\Topc$ in Eq.(\ref{eq:gK}) is also redundant in the history Hilbert space picture, as noted above.

When discretized on the time axis, $\gI[\boldAdop,\boldAop]$ would naturally correspond to the part within the dashed box in Fig.~\ref{fig:qiptn}(a), $\gK[\Himp]$ corresponds to the bare system propagators as indicated by the red squares in Fig.~\ref{fig:qiptn}(a), which are the reasons why the modified trace $\Trpt$ of PT is used in Eq.(\ref{eq:Zimp}).

% To this end, we note that compared to the Feynman-Vernon IF used for diagonal and commutative system-bath coupling, i.e., $\gI[\bolds] = e^{\int_{\contour}\dd \tau \int_{\contour} \dd \tau' s(\tau) \Delta(\tau, \tau') s(\tau')}$ where $s(\tau)$ is an eigenvalue of $\Aop(\tau)$, $\gI[\boldAdop, \boldAop]$ in Eq.(\ref{eq:IF}) is a funtional of operator paths instead of variable paths. In case $\Adop =\Aop$, Eq.(\ref{eq:IF}) reduces to the Feynman-Vernon IF used for diagonal and non-commutative system-bath coupling~\cite{gribben2022-exact,Chen2024}.
Eq.(\ref{eq:IF}) is exact in the continuous-time limit.
Therefore by using it as the starting point of a numerical algorithm, the bath discretization error as well as the local bosonic Hilbert space truncation error could be completely eliminated. 
Next we will show the algorithm we use to construct the IF in Eq.(\ref{eq:IF}) as an MPO, which corresponds to the MPO in Fig.~\ref{fig:qiptn}(b), and this MPO will be referred to as the MPO-IF afterwards. 

To this end, we discuss the possible generalization of Eq.(\ref{eq:IF}) to the case of an arbitrary number of system operators coupled to the same bath, as considered in Ref.~\cite{Link2026}. The present work has derived the IF for a conjugate pair $\Adop$ and $\Aop$ coupled to a Gaussian bosonic bath, and has demonstrated the multi-bath construction where the system couples to several independent baths each via its own conjugate pair, with the full IF obtained by multiplying the individual MPO-IFs. The situation of an arbitrary number $L$ of Hermitian system operators coupled to the same bath is more general than the conjugate-pair case and remains a proposed extension. The Hamiltonian is
\begin{align}
\Hhyb' = \sum_{k} \sum_{\nu=1}^L \Xop_{\nu}(V_{\nu, k}\bdop_k + \hc),
\end{align}
where $\Xop_{\nu}$ are Hermitian operators and $L$ is the total number of $\Xop_{\nu}$. $\Hhyb'$ can be reduced to our case by setting $\Xop_{1}=\Xop$, $\Xop_2 = \Yop$, $V_{1,k} = V_k$ and $V_{2, k} = -\im V_k$.
For this more general $\Hhyb'$, the exponent in Eq.(\ref{eq:IF}) would acquire a dependence on $\nu$, which could likely be generalized into $-\int_{\contour}\int_{\contour} \sum_{\nu,\nu'=1}^L \Xop_{\nu}(\tau) \Delta_{\nu, \nu'} (\tau, \tau')\Xop_{\nu'}(\tau')$ instead, with $\Delta_{\nu, \nu'}(\tau, \tau') = \im \int \dd\omega J_{\nu, \nu'}(\omega) D_{\omega}(\tau, \tau')$ and $J_{\nu, \nu'}(\omega) = \sum_k V_{\nu, k}^{\ast} V_{\nu', k}\delta(\omega-\omega_k)$. However, the rigorous verification of this expression is beyond the scope of this work. Assuming that this is the correct expression in the general case, then the rank of the full tensor obtained by discretizing the IF would only scale linearly with the total number of discrete time steps $N$ and be independent of $L$, which is similar to Eq.(\ref{eq:IF}) and different from Ref.~\cite{Link2026} (the rank of the full tensor in the latter scales linearly with both $L$ and $N$). This scaling statement is conjectural and remains to be verified. In addition, for the specific type of off-diagonal system-bath coupling considered in this work, the number of terms in the general expression for the exponent would be four times that in the exponent of Eq.(\ref{eq:IF}), which is because this more general expression has not taken into account the fact that the two coupling terms in Eq.(\ref{eq:Hhyb2}) are not independent. Therefore our current expression in Eq.(\ref{eq:IF}) is likely to be more efficient for numerical calculations in the considered case, although numerical performance of generalized TEMPO methods will ultimately be determined by the required bond dimension of the underlying MPS/MPO.

% To this end, we discuss the difference between Eq.(\ref{eq:IF}) and the case of diagonal system-bath coupling. In the latter case one could choose the eigenbasis of the Hermitian operator $\Aop$, and then one can rewrite Eq.(\ref{eq:IF}) as a functional of the eigenvalues of $\Aop$ as
% \begin{align}\label{eq:IFdiagonal}
% \gI[\bolds] = e^{\int_{\contour}\dd \tau \int_{\contour} \dd \tau' s(\tau) \Delta(\tau, \tau') s(\tau')},
% \end{align}
% where $s(\tau)$ is an eigenvalue of $\Aop(\tau)$ and $\bolds$ denotes the path of eigenvalue.

\subsection{Algorithm to build the MPO-IF}

\begin{figure}
  \includegraphics[]{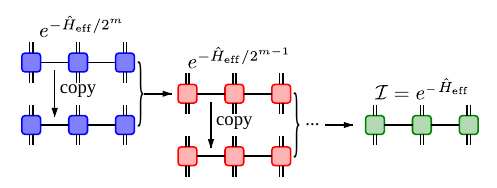} 
  \caption{The exponential tensor renormalization group algorithm to build the Feynman-Vernon IF as an MPO. By choosing a step size $1/2^m$, the effective thermal state $e^{-\Heff}$ can be obtained using only $m$ MPO-MPO multiplications.
    }
    \label{fig:xtrg}
\end{figure}

Similar to QuAPI and TEMPO, the first step of our method is to discretize the operator paths in Eq.(\ref{eq:IF}) using the QuAPI scheme (see Appendix.~\ref{app:quapi} for details), which yields the following finite-$\delta t$ approximation of the contour-ordered IF on the Keldysh contour
\begin{align}\label{eq:disIF}
\gI[\boldAdop, \boldAop] \approx e^{-\sum_{\zeta, \zeta'=\pm}\sum_{i,j=1}^N \Adop_{i,\zeta}\Delta^{\zeta\zeta'}_{i,j}\Aop_{j,\zeta'} },
\end{align}
where $N=t/\delta t$, $+$ means the forward ($0\rightarrow t$) and $-$ means backward ($t\rightarrow 0$) branches on $\contour$ respectively.
The QuAPI discretization is based on a first-order short-time splitting. Its local error is of order $\mathcal{O}(\delta t^2)$, and the resulting global error at fixed final time is expected, in general, to be of order $\mathcal{O}(t\delta t)$. This error is distinct from the Prony-decomposition, XTRG, and MPO-truncation errors and must be checked by convergence with respect to $\delta t$ for the target time interval. 
We can see that $\gI[\boldAdop, \boldAop]$ has the same form as the \textit{thermal state of an effective long-range Hamiltonian}:
\begin{align}\label{eq:Heff}
\Heff = \sum_{\zeta, \zeta'=\pm}\sum_{i,j=1}^N \Adop_{i, \zeta} \Delta^{\zeta\zeta'}_{i,j}\Aop_{j, \zeta'}
\end{align}
with inverse temperature $1$, therefore one could use the well-established 1D TN techniques to build $e^{-\Heff}$ as an MPO~\cite{Schollwock2011}. In this work, we opt to use the  exponential tensor renormalization group (XTRG) algorithm~\cite{ChenWeichselbaum2018} which can converge exponentially fast to the thermal state. We note that this algorithm has already been used for solving fermionic impurity problems~\cite{GuoChen2024d}. 
By choosing a step size $\delta = 1/2^m$ ($\delta$ is completely different from the step size $\delta t$ used to discretize the Keldysh contour), one can first build $e^{-\Heff/2^m}$ as an MPO, which can be done using the $\WI$ or $\WII$ algorithm~\cite{ZaletelPollmann2015} for example. Then one could multiply two such MPOs together to obtain $e^{-\Heff/2^{m-1}}$ as an MPO, and the final $e^{-\Heff}$ can be obtained by repeating this procedure for $m$ times. The XTRG algorithm to build the MPO-IF is schematically illustrated in Fig.~\ref{fig:xtrg}.
% A central operation in this algorithm is the multiplication of two same MPOs with bond dimension $\chi$, which will result in a new MPO with bond dimension $\chi^2$, 
The multiplication of two MPOs with bond dimension $\chi$ will result in an intermediate MPO with bond dimension $\chi^2$, which needs to be compressed back into a new MPO with bond dimension $\chi$ by canonicalizing the intermediate MPO and performing singular value truncation therein. However, the manipulation of the intermediate MPO would be very expensive as the cost would scale as $O(N\chi^6)$ (the singular value decomposition of each tensor in the intermediate MPO has a cost of $O(\chi^6)$). In this situation, the iterative scheme for MPO-MPO multiplication is more advantageous, which first initializes an MPO with bond dimension $\chi$ and then iteratively minimizing its distance to the multiplication of the two input MPOs, whose cost only scales as $O(N\chi^4d^3)$ (the cost of the largest tensor contraction is $O(\chi^4d^3)$ in the iterative MPO-MPO multiplication scheme)~\cite{Schollwock2011}. Still, such a scaling could be numerically very challenging for structural baths with large memory size (large $\chi$), for large impurities (large $d$), or for extremely long-time dynamics (very large $N$).

In our numerical implementation of the iterative MPO-MPO multiplication scheme, we use the single-site version which initializes the output MPO with bond dimension $\chi$ and then updates the tensors site by site using density matrix renormalization group-like sweeps (singular value truncation is not required in single-site updates). We use at most $10$ sweeps, but stop in advance if the standard deviation of the residuals within a sweep is less than $10^{-12}$. As a result, all the MPOs that are actually generated in our algorithm, including the final MPO-IF, will be enforced to have the bond dimension $\chi$.

Once the MPO-IF has been built, one could easily absorb the influence of $\Himp$ into the MPO-IF to obtain the PT as shown in Fig.~\ref{fig:qiptn}(b). With the PT, one could then calculate any multi-time correlation functions of the system as shown in Fig.~\ref{fig:pt}(d,e). In addition, we point out that for QIPs the PT of the system not only encodes all the information of the system dynamics, but essentially encodes the whole information of the system and the bath dynamics, as one could systematically convert any observables of the bath into equivalent multi-time correlations of the system only, by exploiting the special mathematical structure of QIPs~\cite{gribben2021-using,Chen2025}.

Now we discuss the general case that the system is coupled to multiple baths, i.e., $\Hint = \sum_i\Hint^l$ where $\Hint^l$ denotes the coupling to the $l$th bath. In this case, one can simply build an MPO-IF for each $\Hint^l$ independently, since in the picture of the first-order Trotter decomposition, one has $e^{\Lop\delta t}\approx e^{\Limp\delta t}\prod_i e^{\Lint^i\delta t}$, and the IF corresponding to each $\Hint^l$ can be derived independently by tracing out its own bath.
After that, the final MPO-IF which encodes the influence of $\Hint$ can be obtained by multiplying these MPOs together (this multiplication does not have to be actually performed, but can be done on the fly when calculating observables~\cite{ChenGuo2024a}). This procedure is the same as the one used in the case of diagonal and non-commutative system-bath coupling~\cite{gribben2022-exact,Chen2024}. In this approach, the price to pay to deal with $M$ baths is two-fold: (1) one needs to build a separate MPO-IF for each bath, whose cost is simply $M$ times the cost for each bath; (2) for calculating observables, one needs to multiply these MPO-IFs on the fly with a cost $O(N\chi^{M+1})$, compared to $O(N\chi^2)$ for a single bath. A brief theoretical proof is given in Appendix.~\ref{app:multiplebaths} for dealing with multiple baths under the TEMPO framework.

% and then multiply these MPOs together using standard MPO-MPO multiplication to obtain the MPO-IF for the influence of $\Hint$. The reason for this is because in the picture of first-order Trotter decomposition, one has $e^{\Lop\delta t}\approx e^{\Limp\delta t}\prod_l e^{\Lint^l\delta t}$, and the Feynman-Vernon IF for each $\Hint^l$ can be done independently as the baths are independent.

\subsection{Relation to the case of diagonal system-bath coupling}

\begin{figure}
  \includegraphics[]{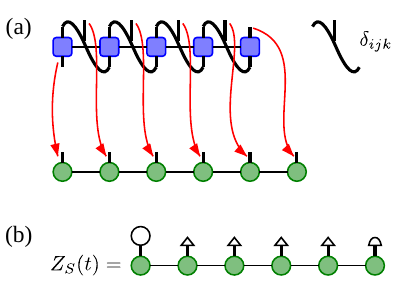} 
  \caption{(a) Converting a PT of $N$ time steps into an ADT of $N+1$ time steps by applying a 3D copy tensor on each pair of neighbouring site tensors. (b) Calculating the partition function based on the ADT, where the boundaries can be treated in the same way as the case of PT, while the middle indices are simply summed over, indicated by the empty triangular hats. Here we have used thick solid lines for the double legs for brevity.
    }
    \label{fig:pt2adt}
\end{figure}

It would be insightful to see how our method reduces to TEMPO and its previous extensions.
First, the case of diagonal and non-commutative system-bath coupling (to multiple baths) is simply a special instance of Eq.(\ref{eq:disIF}) with $\Adop_l=\Aop_l$ for each bath, and our strategy to build the MPO-IF can be directly used.

For the case of diagonal and commutative system-bath coupling, as considered in QuAPI and original TEMPO,
the system path integral is usually written as
\begin{align}
\Zimp(t) = \int\mathcal{D}[\bolds] \gI[\bolds]\gK[\bolds],
\end{align}
where $\gI[\bolds]$ is the Feynman-Vernon IF in terms of the variables path $\bolds = \{s(t)\}$ ($s(t)$ can be viewed as an eigenvalue of the Hermitian operator $\Aop(t)$):
\begin{align}\label{eq:IFdiagonal}
\gI[\bolds] = e^{-\int_{\contour}\dd \tau \int_{\contour} \dd \tau' s(\tau) \Delta(\tau, \tau') s(\tau')},
\end{align}
and $\gK[\bolds]$ can be obtained from Eq.(\ref{eq:gK}) by inserting complete basis of the system at each time step. 
% \begin{align}
% \gK[\bolds] = 
% \end{align}
After discretization, Eq.(\ref{eq:IFdiagonal}) becomes 
\begin{align}\label{eq:disIFnormal}
\gI[\bolds] = e^{-\sum_{\zeta, \zeta'=\pm}\sum_{i,j=1}^N s_{i,\zeta}\Delta^{\zeta\zeta'}_{i,j}s_{j,\zeta'} }.
\end{align}
We can see that Eq.(\ref{eq:disIFnormal}) denotes a tensor $\gI[\bolds]$ of rank-$2N$.
In terms of $\bolds$, $\gK[\bolds]$ is also a tensor of rank-$2N$. Performing element-wise product between $\gK[\bolds]$ and $\gI[\bolds]$, one obtains the integrand of the system path integral as a single tensor, which is conventionally referred to as the augmented density tensor (ADT)~\cite{StrathearnLovett2018}.
In comparison, Eq.(\ref{eq:disIF}) denotes a tensor of rank-$4N$, and the influence of $\Himp$, denoted as $\gK[\Himp]$, is simply a separable set of local operations (Fig.~\ref{fig:qiptn}), which can be absorbed into $\gI[\Adop, \Aop]$ to obtain the PT. Therefore, we can see that the important differences between our method and QuAPI/TEMPO are (i) in our method an elementary operation between tensors is the standard tensor product, while in the latter it is the element-wise product, and (ii) in TEMPO the tensors are naturally represented as MPSs, while in our method the tensors are naturally represented as MPOs.
% Eq.(\ref{eq:disIFnormal}) denotes a tensor of rank-$2N$, which is conventionally referred to as the augmented density tensor (ADT), while in comparison Eq.(\ref{eq:disIF}) denotes a tensor of rank-$4N$, which is a PT up to local operations. 
Meanwhile, the PT can also be systematically converted into the ADT by applying a 3D copy tensor between each neighbouring pair of site tensors, as shown in Fig.~\ref{fig:pt2adt}, where the 3D copy tensor is defined as $\delta_{ijk}=1$ for $i=j=k$, and $0$ otherwise.

The exponent in Eq.(\ref{eq:disIFnormal}) can be viewed as a ``classical Hamiltonian'' as all its terms commute with each other ($\gI[\bolds] $ can thus be viewed as the partition function of a classical statistics model). Exploiting this feature, one can rewrite $\gI[\bolds]$ as 
\begin{align}\label{eq:partialif}
  \gI[\bolds] = \prod_{i, \zeta} e^{-\sum_{j, \zeta'}s_{i,\zeta}\Delta^{\zeta\zeta'}_{i,j}s_{j,\zeta'}},
\end{align}
where each $e^{-\sum_{j, \zeta'}s_{i,\zeta}\Delta^{\zeta\zeta'}_{i,j}s_{j,\zeta'}}$ is a \textit{partial IF}. Importantly, it was shown that each partial IF can be compactly built as an MPS with bond dimension $d$ only~\cite{StrathearnLovett2018,GuoChen2024d}, and the MPS representation of $\gI[\bolds] $ can thus be built by multiplying $2N$ such partial IFs together. The partial IF algorithm can also be straightforwardly applied to the case of diagonal and non-commutative system-bath coupling, as the commutation relation of the effective Hamiltonian in these two cases are exactly the same. 

However, for off-diagonal system-bath coupling, the terms in Eq.(\ref{eq:Heff}) do not commute in general, therefore the partial IF algorithm cannot be used.
In fact, the strategy used in this work is a direct generalization of the \textit{time-translationally invariant IF} algorithm proposed in Ref.~\cite{GuoChen2024d}, which can be applied even if $\Heff$ is a ``quantum Hamiltonian''. A similar strategy has also been applied to the case of diagonal system-bath coupling~\cite{YangFang2026}, but the time-dependent variational principle (TDVP) algorithm~\cite{HaegemanVerstraete2016} is used to build the effective thermal state, instead of the XTRG algorithm used in this work.
An additional advantage of this strategy is that it explicitly preserves the time-translational invariance (TTI) of the IF on the time axis (essentially because $\Hint$ is time-independent), which can thus be directly used in combination with infinite MPS techniques, as has already been explored in the fermionic case~\cite{GuoChen2024e,GuoChen2024d}. In comparison, the TTI property is broken in the partial IF algorithm (we note the recent works which restore the TTI property with a variant of the partial IF algorithm~\cite{LinkStrunz2024,GarbelliniStrunz2026}).

\subsection{Possible generalization to the fermionic case}
Here we briefly address the possible generalization of our method to the fermionic case, and its relation to existing approaches therein. The system-bath coupling in the fermionic case is off-diagonal in general, with the form $\Hhyb = \sum_k(V_k\ddop\cop_k + \hc)$ if only a single fermionic flavor is considered, where $\ddop$ and $\dop$, $\cdop_k$ and $\cop_k$ are the fermionic creation and annihilation operators of the system and bath respectively. 
% The analytic expression of the Feynman-Vernon IF is (see Ref.~\cite{GullWerner2011} for example)
% \begin{align}\label{eq:disIFfermion}
% \gI[\boldddop, \bolddop] = e^{\int_{\contour} \dd\tau\int_{\contour}\dd \tau' \ddop(\tau)\Delta(\tau, \tau')\dop(\tau')},
% \end{align}
% in correspondence with Eq.(\ref{eq:IF}) and Eq.(\ref{eq:IFdiagonal}). 
Drawing correspondence to Eq.(\ref{eq:IF}), one may write the analytic expression of the Feynman-Vernon IF in the fermionic case as (we note that the following expression may not be exact as the IF for the fermionic case is generally derived in the coherent state basis and written in terms of Grassmann variables):
\begin{align}\label{eq:disIFfermion}
\gI[\boldddop, \bolddop] = e^{-\int_{\contour} \dd\tau\int_{\contour}\dd \tau' \ddop(\tau)\Delta(\tau, \tau')\dop(\tau')}.
\end{align}
If this expression is valid, then similar to the bosonic case, one could use the QuAPI scheme to discretize the IF~\cite{ChenGuo2024a}, after which the exponent of the discretized IF will be an \textit{effective fermionic Hamiltonian}, and one could build it as a fermionic MPO using standard 1D fermionic TN techniques~\cite{BultinckVerstraete2017}. 

However, this approach has not been taken so far in current developments of fermionic analogies to the TEMPO method (which might be worth to be explored in the future). Instead, two different approaches have been proposed. The first approach, proposed in Refs.~\cite{ThoennissAbanin2023b,NgReichman2023}, used an idea similar to the fermionic thermofield mapping~\cite{BlasoneVitiello2011} to map the effective thermal state into an equivalent pure state in an enlarged Fock state basis (which essentially squashes a density matrix into a vector). The correspondence to the PT would then become obscure in this approach. In the second approach, one uses the usual Feynman-Vernon IF for fermions in terms of Grassmann variables, and then represents the IF as a Grassmann MPS~\cite{ChenGuo2024a}. As such the second approach, referred to as the Grassmann TEMPO (GTEMPO) method, essentially builds the MPO-IF in the coherent state basis. Importantly, in the coherent state basis, all the terms in $\Heff$ commute with each other due to the Grassmann algebra and the quadratic form of $\Heff$. As a result, $\Heff$ becomes similar to a classical Hamiltonian, and the techniques used in original TEMPO, such as the partial IF algorithm, can be straightforwardly adopted in GTEMPO.

To this end, we note an important feature that is shared by QuAPI/TEMPO and our extended TEMPO (and by GTEMPO in the fermionic case): their core algorithms are self-contained and only depend on the discretized expression of the Feynman-Vernon IF. As a result, the problem can be solved by giving $\Himp$ and the hybridization function $\Delta(\tau, \tau')$ only, without referring to further details of the system-bath coupling. This feature is the same as the continuous-time quantum Monte Carlo methods~\cite{GullWerner2011}, which could be very beneficial for studying phenomenological system-bath couplings or in BDMFT.

% \section{Brief review of the process tensor framework}

% \section{Process tensor representation of the Feynman-Vernon IF}

\section{Numerical results}\label{sec:results}

There are four error sources in our method in total: (1) the time discretization error characterized by the real-time step size $\delta t$ (or the imaginary-time step size $\delta\tau$ if imaginary-time evolution is considered); (2) the MPO bond truncation error characterized by $\chi$; (3) the error in approximating the hybridization function as the sum of exponential functions, which is used to build the efficient MPO representation of $\Heff$ (see Appendix.~\ref{app:xtrg} for details), characterized by the number of exponential functions $n$; (4) another ``time discretization error'' (with a time step size $\delta=1/2^m$) occurred in the XTRG algorithm to build the effective thermal state $e^{-\Heff}$, characterized by an integer $m$. In addition, if the system itself is a bosonic mode, then we need to choose a cutoff for the local Hilbert space, still denoted as $d$, which will also affect the numerical accuracy.
In our numerical examples we will analyze the effects of all these hyperparameters.
We note that it has been shown that for BSFs satisfying a moderate continuity requirement, $n$ grows only logarithmically as $t$, and thus one would expect that a small $n$ would suffice in general~\cite{VilkoviskiyAbanin2024,ThoennissAbanin2024,HuangLin2026b}. In the meantime, a defining feature of the XTRG algorithm is that it converges exponentially fast with $m$~\cite{ChenWeichselbaum2018}. Therefore we expect that the numerical results would in general quickly saturate with a small $m$ and $n$. In our numerical simulations we will set $m=7$ and $n=20$ by default, unless we particularly analyze the numerical convergence against them.

In the following, we divide our numerical simulations into two categories. In the first category, we validate our method against exactly solvable cases such that we have a perfect benchmarking baseline, including toy models in which the baths contain only a single bosonic mode and the noninteracting case where the impurity itself is a noninteracting bosonic mode (in the latter case the total Hamiltonian is quadratic and the problem can be efficiently and exactly solved). In the second category, we apply our method to study nonintegrable cases on both the Keldysh and imaginary contours, to illustrate the effectiveness and flexibility of our method.

% In our numerical simulations, we consider the Jaynes-Cummings-type system-bath coupling to a single bath, with $\Hhyb$ in Eq.(\ref{eq:jccoupling}). The more general case of multiple baths can be straightforwardly solved by constructing an MPO-IF for the coupling to each bath and multiply these MPO-IFs together as discussed in Sec.~\ref{sec:method}, which will not be numerically studied in this work.

% In the following, we will first validate our method in two exactly solvable cases: the JC model where the spin is coupled to a bath that contains a single bosonic mode, and the free bosons model where the system itself is a noninteracting bosonic mode. Then we consider the more general case that the spin is coupled to a subohmic bath via JC-type coupling, which is often used to approximate the standard spin-boson model by neglecting the contour-rotating terms (i.e., the rotating-wave approximation). We will compare the results against those from solving the standard spin-boson model using TEMPO, for coupling strengths from weak to strong.

\subsection{Validation of our method}

\subsubsection{The Jaynes-Cummings model}
\begin{figure}
  \includegraphics[]{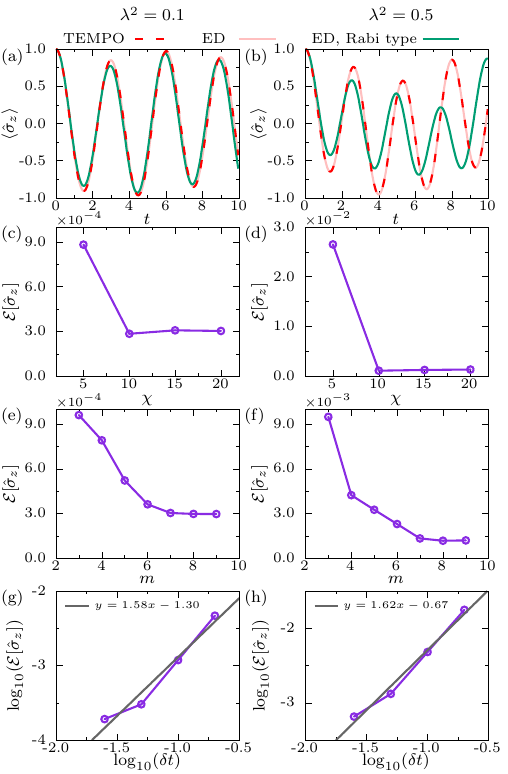}
  \caption{(a,b) Real-time evolution of the average spin $\langle\sgz\rangle$ of the JC model with (a) $\lambda^2=0.1$ and (b) $\lambda^2=0.5$. The red dashed lines are extended TEMPO results calculated using $\chi=30$ and $\delta t=0.05$, and the red solid lines are the corresponding ED results. The cyan solid lines are the ED results for the corresponding Rabi model. We have used a local Hilbert space cutoff $d_c=50$ for the single-mode bath in these ED calculations. (c,e,g) The mean error $\mathcal{E}[\sgz]$ of $\langle\sgz\rangle$ between the extended TEMPO results and the ED results for $\lambda^2=0.1$ as a function of $\chi$ (c), $m$ (e) and $\delta t$ (g) respectively. (d,f,h) The same plots as in (c,e,g) but for $\lambda^2=0.5$. In panels (c-g) we have used the default parameters $\chi=30$ and $\delta t=0.05$ unless they are particularly analyzed against. 
  We have used log scales for both axes in (g,h), where the purple circles show the four data points with $\delta t=0.025, 0.05, 0.1, 0.2$ respectively, and the gray solid line is the polynomial fitting. 
  We have chosen $\rhoimp=\vert e\rangle\langle e\vert$ as the system initial state, and used $\Omega=\omega_0=1$ and $\beta=5$ in all these simulations. 
    }
    \label{fig:jcmodel}
\end{figure}

The Hamiltonian of the Jaynes-Cummings (JC) model can be written as
\begin{align}\label{eq:JCmodel}
\Hop = \Omega\sgz + \lambda (\sgp\bop+\sgm\bdop) + \omega_0\bdop\bop,
\end{align}
where $\sgx, \sgy, \sgz$ are Pauli operators, $\hat{\sigma}_{\pm} = (\sgx\pm\im\sgy)/2$, $2\Omega$ is the energy gap of the spin, $\omega_0$ is the frequency of the bosonic mode, and $\lambda$ is the coupling strength. The JC model can be viewed as a very special QIP in which the spin is coupled to a single-mode bath. The BSF is simply $J(\omega) = \lambda^2\delta(\omega - \omega_0)$ in this case, the hybridization function in Eq.(\ref{eq:hybridization}) is exactly a single exponential function and the hyperparameter $n=1$ will be sufficient.
The JC model is often used to approximate the Rabi model with coupling $\Hhyb=\lambda\sgx(\bop+\bdop)$, by employing the rotating-wave approximation (RWA) which neglects the counter-rotating term $\lambda(\sgp\bdop+\sgm\bop)$ in the Rabi model for small $\lambda$.
In our numerical simulations of the JC model, we will choose $\beta=5$ for the bath, and choose $\rhoimp=\vert e\rangle\langle e\rangle$ (i.e., the spin is initially in the excited state). We note that the single-mode bath condition does not formally simplify our extended TEMPO calculations.

In Fig.~\ref{fig:jcmodel}(a,b), we plot the average spin, defined as $\langle\sgz\rangle = \tr(\sgz\rhoop) /\tr(\rhoop) $, as a function of time $t$, and compare our extended TEMPO results against the exact diagonalization (ED) results (in which the total Hamiltonian is directly diagonalized) for $\lambda^2=0.1$ (weak coupling) and $\lambda^2=0.5$ (strong coupling) respectively. We have considered $\beta=5$ in these simulations, and used $\chi=30$, $\delta t=0.05$ for our extended TEMPO calculations. The ED results of the corresponding Rabi model are also plotted for comparison. We can see that for small $\lambda$, the JC model is indeed a good approximation of the Rabi model, and that for both coupling strengths, our extended TEMPO results agree very well with the ED results.

For the single-mode bath, the hybridization function is a single exponential function, therefore $\Heff$ can be exactly built as an MPO (see Appendix.~\ref{app:xtrg}) and the hyperparameter $n$ does not come into play, which makes it an ideal test ground for error analysis against the remaining hyperparameters.
In Fig.~\ref{fig:jcmodel}(c,e,g), we analyze the mean error $\mathcal{E}$, defined as $\mathcal{E} = \sqrt{||\vec{x} -\vec{y}||^2/L}$ for two vectors $\vec{x}, \vec{y}$ of length $L$, between the extended TEMPO results and the ED results for $\lambda^2=0.1$ as functions of $\chi$ (c), $m$ (e) and $\delta t$ (g) respectively. In Fig.~\ref{fig:jcmodel}(d,f,h), we perform the same error analysis as in Fig.~\ref{fig:jcmodel}(c,e,g), but for $\lambda^2=0.5$ instead. 
% The error against $n$ is not considered here since $n=1$ is already exact for single-mode bath. 
For both coupling strengths, we can see that the error quickly saturates at $\chi=10$ and $m=7$, and that the error decreases monotonically for smaller $\delta t$ with a faster-than-linear behavior (the rate is going to be smaller if we further decrease $\delta t$, as already indicated by the point with smallest $\delta t=0.025$). The errors for $\lambda^2=0.5$ are generally more than one order of magnitude larger than those for $\lambda^2=0.1$ under the same hyperparameters, which indicates that the strong coupling regime could be harder to simulate for our method.
% Interestingly, from the definition of the PT in Fig.~\ref{fig:pt}(a), the bond dimension of the PT for a single bosonic mode will be $\chi=d_c^2$ if we truncate the local bosonic Hilbert space to $d_c$, while our extended TEMPO calculations already obtain very accurate results ($\mathcal{E}<10^{-3}$) with $\chi=10$ only, which means that even for a single-mode bath where the hybridization function is purely oscillating, the information in the PT can still be significantly compressed.

\subsubsection{A spin coupled to two single-mode baths}
\begin{figure}
  \includegraphics[]{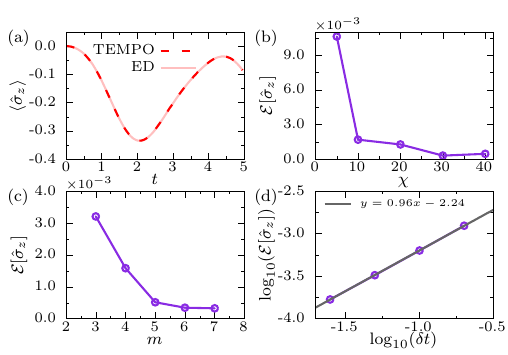}
  \caption{(a) Real-time evolution of the average spin $\langle\sgz\rangle$ of the toy model defined in Eq.(\ref{eq:toymodel}), where the red dashed line represents the extended TEMPO results calculated using $\chi=30$ and $\delta t=0.05$, and the red solid line represents the corresponding ED results calculated using $d_c=25$ for the bath mode. (b, c, d) The mean error $\mathcal{E}[\sgz]$ between the extended TEMPO results and the ED results as a function of $\chi$ (b), $m$ (c) and $\delta t$ (d) respectively, where we have used the default hyperparameters $\chi=30$ and $\delta t=0.05$ unless they are particularly analyzed against. 
  We have used log scales for both axes in (d), where the purple circles show the four data points with $\delta t=0.025, 0.05, 0.1, 0.2$ respectively, and the gray solid line is the polynomial fitting. 
  We have chosen $\rhoimp = (\vert g\rangle\langle g\vert + \vert e\rangle\langle e\vert)/2$ as the system initial state, and used $\lambda_l^2=\lambda_r^2=0.1$, $\Omega=\omega_l=\omega_r=1$, $\beta_l=1$, $\beta_r=10$ in all these simulations.
    }
    \label{fig:twobaths}
\end{figure}

Now we consider a toy model in which a spin is simultaneously coupled to two single-mode baths, to illustrate the ability of our method to deal with non-commutative and off-diagonal coupling to multiple baths. The total Hamiltonian is chosen as:
\begin{align}\label{eq:toymodel}
\Hop = &\Omega\sgz + \lambda_l (\sgp\bop_l+\sgm\bdop_l) + \omega_l\bdop_l\bop_l \nonumber \\ 
&+ \lambda_r\sgx(\bdop_r + \bop_r) + \omega_r\bdop_r\bop_r,
\end{align}
where the spin is off-diagonally coupled to the left single-mode bath defined by $\bdop_l$, $\bop_l$ with strength $\lambda_l$, and diagonally coupled to the right single-mode bath defined by $\bdop_r$, $\bop_r$ with strength $\lambda_r$. In our numerical simulations of this model, we choose $\lambda_l^2=\lambda_r^2=0.1$ and $\Omega = \omega_l=\omega_r=1$. The inverse temperatures of the initial thermal states of the two baths are chosen to be $\beta_l=1$ and $\beta_r=10$ for the left and right baths respectively. The initial state of the spin is chosen to be the infinite-temperature state, i.e., $\rhoimp = (\vert g\rangle\langle g\vert + \vert e\rangle\langle e\vert)/2$ where $g$ and $e$ denote the ground and excited states of the spin.
As has been discussed in Sec.~\ref{sec:method}, in this case we can simply build an MPO-IF for the coupling to each bath independently and then multiply the resulting MPOs together to obtain the MPO-IF which represents the effects of all baths.

The results are plotted in Fig.~\ref{fig:twobaths}. In Fig.~\ref{fig:twobaths}(a), we plot the time evolution of the average spin as a function of real time $t$, and compare our extended TEMPO results calculated using $\chi=30$ and $\delta t=0.05$ to the ED results, where we can see a very good match between the two sets of results. In Fig.~\ref{fig:twobaths}(b,c,d), we analyze in detail the mean error $\mathcal{E}$ of the average spin between the extended TEMPO results and the ED results  as functions of $\chi$ (b), $m$ (c) and $\delta t$ (d) respectively. Similar to the case of a single bath, we can see that the mean error decreases with refining hyperparameters. From Fig.~\ref{fig:twobaths}(b,c), we can see that $\mathcal{E}$ saturates at around $\chi=30$, $m=6$ for fixed $\delta t=0.05$. From Fig.~\ref{fig:twobaths}(d), we can see that $\mathcal{E}$ decreases linearly with smaller $\delta t$, which indicates that under these parameters the time discretization is the dominating error source.

\subsubsection{The noninteracting case}

\begin{figure}
  \includegraphics[]{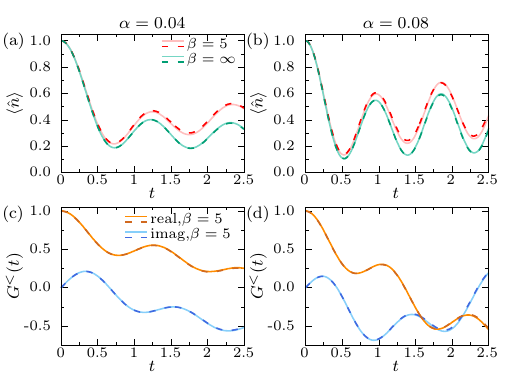}
  \caption{(a,b) The average occupation $\langle \nop\rangle $ as a function of $t$ for $\alpha=0.04$ (a) and $\alpha=0.08$ (b). The red and cyan dashed lines are extended TEMPO results for $\beta=5$ and $\beta=\infty$ respectively. The solid lines with the same colors are the corresponding ED results. (c, d) The real (orange lines) and imaginary (blue lines) parts of the lesser Green's function $G^<(t)$ as a function of $t$ at $\beta=5$ for $\alpha=0.04$ (c) and for $\alpha=0.08$ (d), where the dashed and solid lines with the same color are the extended TEMPO results and the corresponding ED results. We have used $\chi=30$, $\delta t=0.00625$ in these extended TEMPO simulations, and used $\delta\omega=0.01$ in the ED simulations.
  %(e, f) The mean error for the average occupation between our extended TEMPO results and ED results as a function of $\delta t$ (e), and of $\chi$ (f). In (a,b,c,d,e) we have fixed $\chi=80$ for our extended TEMPO calculations. In (f) we have fixed $\delta t=0.00625$. We have chosen $\rhoimp=\vert 1\rangle\langle 1\vert$ and $d=4$ in these simulations.
    }
    \label{fig:freebosons}
\end{figure}

\begin{figure}
  \includegraphics[width=\columnwidth]{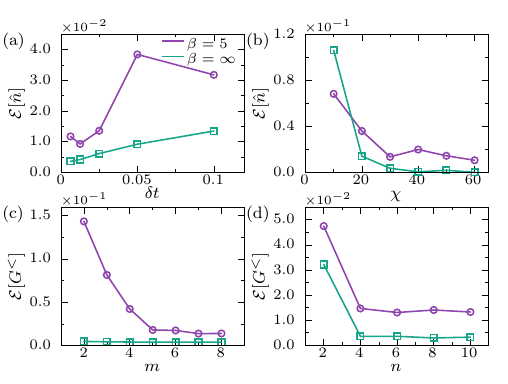} 
  \caption{The mean error of the average occupation between the extended TEMPO results and the corresponding ED results for the noninteracting case with $\alpha=0.08$, against the four hyperparameters $\delta t$ (a), $\chi$ (b), $m$ (c) and $n$ (d). The purple lines with circle and the cyan lines with square represent the results for $\beta=5$ and $\beta=\infty$ respectively. We have chosen the default hyperparameters as $\chi=30$ and $\delta t=0.025$ for $\beta=5$, and as $\chi=30$ and $\delta t=0.00625$ for $\beta=\infty$. We have also used $d=4$ for $\beta=5$ and $d=2$ for $\beta=\infty$ in these simulations.
    }
    \label{fig:freebosons_error}
\end{figure}

Next, we will consider a continuous subohmic bath with the BSF
\begin{align}\label{eq:subohmic}
J(\omega) = 2\pi\alpha\omega_c^{1-s}\omega^s \Theta(\omega)\Theta(\omega_c-\omega),
\end{align}
which is commonly used in studying the quantum phase transition of the spin-boson model~\cite{ChinPlenio2011,ZhouSun2026}. We will fix $\omega_c=5$ and $s=0.5$ in our numerical simulations. 
% and focus on the case where the bath is initially in the vacuum state (i.e., $\beta=\infty$). 

We first study the noninteracting case where the system itself is a noninteracting bosonic mode, and choose $\Himp =  \adop\aop$ with $\adop, \aop$ the bosonic creation and annihilation operators of the system. The occupation operator of the system is denoted as $\nop = \adop\aop$. 
We will consider two temperatures for the bath: (i) a zero-temperature case with $\beta=\infty$ and (ii) a finite-temperature case with $\beta=5$. For each temperature, we will consider two coupling strengths with $\alpha=0.04$ and $\alpha=0.08$.
The initial state of the system is chosen to be a pure state $\vert 1\rangle$, i.e., the Fock state with a single particle. The noninteracting case can be efficiently solved by diagonalizing a matrix of size $M+1$ if one discretizes the bath into $M$ discrete modes (the bath discretization error is the only source of error in this ED algorithm, which is completely different from the ED algorithm used to solve the JC model), this is possible because the total Hamiltonian is quadratic and the underlying quantum state is a Gaussian state~\cite{GuoPoletti2017,GuoChen2026b}. In our ED calculations for this model, we choose an equal-distant frequency step size $\delta \omega=0.01$ and we have verified that our ED results have well converged against $\delta \omega$. 
In our extended TEMPO simulation of this model, we need to truncate the local Hilbert space of the impurity boson.
In the finite-temperature case with $\beta=5$, we run the extended TEMPO simulations for $d=2,3,4$ until we find that the extended TEMPO results for $d=4$ already match well with the ED results. In the zero-temperature case, the off-diagonal system-bath coupling together with our particular choice of the system-bath initial state allow at most one excitation in the system and bath, therefore $d=2$ suffices.

% \begin{figure}
%   \includegraphics[width=\columnwidth]{figs/freeboson_lesser.pdf} 
%   \caption{XXX.
%     }
%     \label{fig:freebosongf}
% \end{figure}

In Fig.~\ref{fig:freebosons}(a, b), we plot the average occupation $\langle \nop\rangle $ as a function of $t$ for $\alpha=0.04$ (a) and $\alpha=0.08$ (b) respectively. The red and cyan dashed lines are extended TEMPO results for $\beta=5$ and $\beta=\infty$ respectively, calculated using $\chi=30$, $\delta t=0.00625$. The solid lines with the same colors are the corresponding ED results.
In Fig.~\ref{fig:freebosons}(c, d), we plot the lesser Green's function $G^<(t)=\langle\adop(t)\aop\rangle$ as a function of $t$ for $\alpha=0.04$ (c) and for $\alpha=0.08$ (d) respectively, similar to Fig.~\ref{fig:freebosons}(a, b). Here we have only plotted $G^<(t)$ for $\beta=5$, as it is independent of $\beta$ due to the noninteracting condition plus the separable system-bath initial state condition.
We can see that our extended TEMPO results well match the ED results for all these observables.
% From these four panels, we can see that our extended TEMPO results becomes closer to the ED results for smaller $\delta t$, as expected.

In Fig.~\ref{fig:freebosons_error}, we perform error analysis of the extended TEMPO results against the four hyperparameters $\delta t$ (a), $\chi$ (b), $m$ (c) and $n$ (d) for both values of $\beta$, where we have focused on the observable $\nop$ and $\alpha=0.08$. 
The default hyperparameters are chosen as $\chi=30$ and $\delta t=0.025$ for $\beta=5$, and as $\chi=30$ and $\delta t=0.00625$ for $\beta=\infty$ in these simulations.
Generally, we can see that $\mathcal{E}$ converges faster against the hyperparameters for $\beta=\infty$ than for $\beta=5$. In particular, from Fig.~\ref{fig:freebosons_error}(c) the error against $m$ quickly saturates at $m=2$ for $\beta=\infty$, but only saturates at $m=5$ for $\beta=5$. This is likely because the zero-temperature case is a very special case with only one excitation, and our extended TEMPO method could converge more easily as the underlying quantum state can only be weakly entangled. 
The errors tend to decrease with refining hyperparameters as in the previous toy models. However, there are unusual visible jumps of $\mathcal{E}$ against $\delta t$ and $\chi$ for $\beta=5$, which are coincidences that may be attributed to the complex interplay of different error sources. 
For a continuous bath the BSF cannot be described by a single exponential in general, as such we would expect a nontrivial dependence of the error on $n$, which can be seen from Fig.~\ref{fig:freebosons_error}(d). 
Nevertheless, $\mathcal{E}$ saturates very quickly at $n=4$ for both temperatures, as expected.

% we plot the mean error $\mathcal{E}_{\bar{n}}$, defined as $\mathcal{E} = \sqrt{||\vec{x} -\vec{y}||^2/L}$ for two vectors $\vec{x}, \vec{y}$ of length $L$, between $\bar{n}(t)$ calculated by our extend TEMPO and by ED, as a function of $\delta t$ with fixed $\chi=80$. 
% Here we can clearly see that $\mathcal{E}_{\bar{n}}$ is smaller for smaller $\delta t$. In Fig.~\ref{fig:freebosons}(f) we plot $\mathcal{E}_{\bar{n}}$ as a function of $\chi$ for fixed $\delta t=0.00625$, from which we can see that the extended TEMPO results only varies slightly when increasing $\chi$ from $30$ to $80$, indicating that we can already obtain accurate numerical results with extended TEMPO even with a small $\chi=30$ for this problem.

% \begin{figure}
%   \includegraphics[width=\columnwidth]{figs/freeboson_errors.pdf} 
%   \caption{The mean square error between TEMPO results and the ED results as a function of $\delta t$ (a), and $\chi$ (b). 
%     }
%     \label{fig:freebosonerror}
% \end{figure}

% In Fig.~\ref{fig:freebosonerror}, we plot the mean square error between our TEMPO results and ED results as a function of $\delta t$ (a) and $\chi$ (b). 

\subsection{Applications for nonintegrable cases}

\subsubsection{The JC spin-boson model}

% \begin{figure}
%   \includegraphics[]{figs/sb_comparison/fig.pdf} 
%   \caption{The expectation value $\langle \sgz\rangle$ as a function of $t$ for the real-time evolution of the JC spin-boson model where the spin is coupled to a subohmic bath via the JC-type coupling (the solid lines), and of the standard spin-boson model (the dashed lines) under the same parameter settings. The results for the JC spin-boson model are calculated using our extended TEMPO with $\chi=80$, while the results for the spin-boson model are calculated using TEMPO with $\chi=200$. We have chosen $\rhoimp=\vert e\rangle\langle e\vert$, $\beta=\infty$ and $\delta t=0.0125$ in these simulations.
%   % The real-time evolution of the spin-boson model with JC-type coupling (the dashed lines) and with Rabi-type coupling (the solid lines) for different values of $\alpha$. 
%     }
%     \label{fig:comparison}
% \end{figure}

\begin{figure}
  \includegraphics[]{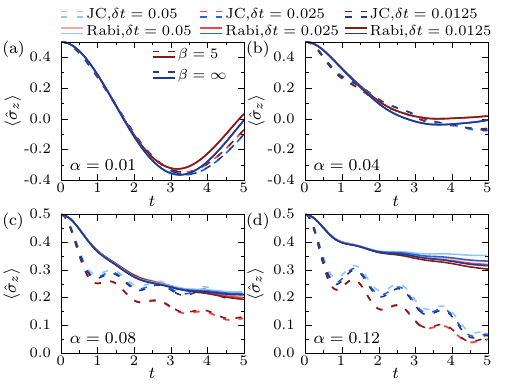}
  \caption{Convergence study of our extended TEMPO results for the JC spin-boson model (the dashed lines) and the TEMPO results for the spin-boson model (the solid lines) against the time step size $\delta t$ for (a) $\alpha=0.01$, (b) $\alpha=0.04$, (c) $\alpha=0.08$, (d) $\alpha=0.12$. The red dashed lines from lighter to darker are extended TEMPO results calculated using $\beta=5$ and $\delta t=0.05, 0.025, 0.0125$ respectively, with fixed $\chi=80$. The red solid lines from lighter to darker are TEMPO results calculated using $\beta=5$ and $\delta t=0.05, 0.025, 0.0125$ respectively, with fixed $\chi=200$. The blue dashed and solid lines represent similar results to the red lines, but for $\beta=\infty$ instead.
    }
    \label{fig:sberrordt}
\end{figure}

\begin{figure}
  \includegraphics[]{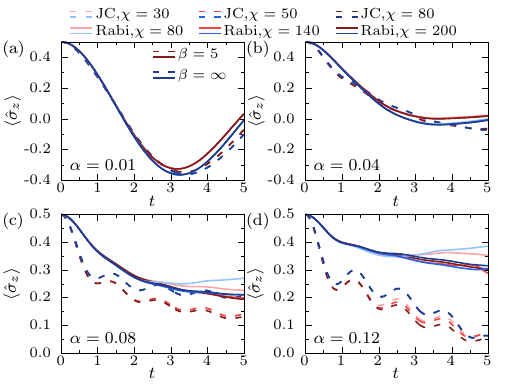}
  \caption{Convergence study of our extended TEMPO results for the JC spin-boson model (the dashed lines) and the TEMPO results for the spin-boson model (the solid lines) against the bond dimension $\chi$ for (a) $\alpha=0.01$, (b) $\alpha=0.04$, (c) $\alpha=0.08$, (d) $\alpha=0.12$. The red dashed lines from lighter to darker are extended TEMPO results calculated using $\beta=5$ and $\chi =30, 50, 80$ respectively. The red solid lines from lighter to darker are TEMPO results calculated using $\beta=5$ and $\chi=80, 120, 200$ respectively. The blue dashed and solid lines represent similar results to the red lines, but for $\beta=\infty$ instead. In these simulations we have fixed $\delta t=0.0125$.
    }
    \label{fig:sberrorchi}
\end{figure}

Now we consider that case of a spin with Hamiltonian $\Himp=\frac{1}{2}\sgz$ that is coupled to a subohmic bath in Eq.(\ref{eq:subohmic}) via JC-type system-bath coupling in Eq.(\ref{eq:jccoupling}) with $\Aop = \frac{1}{2}\sgm$, which will be referred to as the JC spin-boson model in the following. This model can be seen as an approximation of the standard spin-boson model with $\Aop=\frac{1}{2}\sgx$ by neglecting all the counter-rotating terms in $\Hhyb$. It can also be seen as an approximation of a QIP where the system is an interacting bosonic mode and is coupled to a bosonic bath in a way that conserves the total particle number (which is the central problem to solve in BDMFT), and the excitation within the system is so low such that the local bosonic Hilbert space can be truncated to $2$ only.
In addition, this model itself is equivalent to the Lee model in quantum field theory~\cite{Pfeifer1982,LeggettZwerger1987}.
% Similar to the spin-boson model, the JC spin-boson model is nonintegrable in general.
In our numerical simulations, we choose $\rhoimp=\vert e\rangle\langle e\rangle$ and consider two cases with $\beta=\infty$ and $\beta=5$. We note that this model is nonintegrable for general $\beta$, but for the special case with $\beta=\infty$ there is only a single excitation in the model similar to the noninteracting case considered above, and the model becomes analytically solvable. In fact, this is a main reason why the rotating-wave approximation is often employed for the spin-boson model at zero temperature: in this case one could easily obtain a good analytic understanding of the JC spin-boson model, while the original spin-boson model is still nonintegrable~\cite{GuoJohansson2020}. We note that the zero-temperature scenario does not directly simplify our extended TEMPO calculations.

% In Fig.~\ref{fig:comparison}, we study the real-time evolution of the JC spin-boson model and plot the average occupation $\bar{\sigma_z}=\langle\sgz\rangle$ as a function of $t$, for different coupling strengths $\alpha=0.01, 0.04, 0.08, 0.12$. The results for the corresponding spin-boson model are also plotted for comparison, which are calculated using standard TEMPO. We observe revivals of $\bar{\sigma_z}$ for $\alpha=0.01$ for both models, and oscillations at strong couplings at $\alpha=0.08, 0.12$ for the JC spin-boson model. These are purely non-Markovian effects that would be missing if the Born-Markov approximation is employed~\cite{breuer2007-the}. The oscillations for the JC spin-boson model occurring at strong coupling are missing in the spin-boson model, which illustrates that the physics for these two models could be very different for strong coupling. Moreover, even for $\alpha=0.01$, the results for these two models differ significantly, especially for $t>3$, in sharp contrast with the good match in Fig.~\ref{fig:jcmodel}(a),  which indicates that the accuracy of the rotating-wave approximation could be very poor in presence of a structural bath, even in the weak coupling regime.

In Fig.~\ref{fig:sberrordt} and Fig.~\ref{fig:sberrorchi}, we perform convergence study of extended TEMPO calculations against the two hyperparameters $\delta t$ and $\chi$, respectively. The convergence analysis against $m$ and $n$ are not performed in this case, but instead their default values are used, as the bath considered here is the same as the one considered in the noninteracting case and we would expect a similar behavior of errors in the generated MPO-IFs. 
Two sets of calculations are performed in each figure: the red and blue dashed lines are the extended TEMPO results for the JC spin-boson model for $\beta=5$ and $\beta=\infty$ respectively, while the solid lines with the same colors are the corresponding results for the standard spin-boson model calculated using standard TEMPO. 
The four panels in each figure are results for $\alpha=0.01, 0.04, 0.08, 0.12$ respectively. The lines with the same color and the same style (i.e., solid or dashed), but only differ in the gray scale, are results calculated using different values of $\delta t$ in Fig.~\ref{fig:sberrordt} and using different values of $\chi$ in Fig.~\ref{fig:sberrorchi}.

From these two figures, we observe revivals of $\langle \sgz\rangle$ for $\alpha=0.01$ for both models, and oscillations at stronger couplings with $\alpha=0.08, 0.12$ for the JC spin-boson model. These are purely non-Markovian effects that would be missing if the Born-Markov approximation is employed~\cite{breuer2007-the}. The oscillations for the JC spin-boson model occurring at strong coupling are missing in the spin-boson model, which illustrates that the physics for these two models could be very different for strong coupling. Moreover, even for $\alpha=0.01$, the results for these two models still have a visible difference, especially for $t>3$, in sharp contrast with the good match in Fig.~\ref{fig:jcmodel}(a),  which indicates that the accuracy of the rotating-wave approximation could be very poor for the considered structural bath, even in the weak coupling regime.

Generally, we observe that these two sets of calculations can easily converge against $\delta t$ and $\chi$ for small $\alpha$, and that
the extended TEMPO calculations for the JC spin-boson model converge faster than the TEMPO calculations for the spin-boson model as $\delta t$ becomes smaller or $\chi$ becomes larger. 
In particular, for the JC spin-boson model, the extended TEMPO results already converge fairly well at $\chi=50$ and $\delta t=0.025$ for all considered values of $\alpha$, while for the spin-boson model, the TEMPO results have required $\chi=200$ and $\delta t=0.025$ to converge for $\alpha\leq 0.08$, and have not fully converged at $\alpha =0.12$ even when decreasing $\delta t$ from $0.025$ to $0.0125$ or increasing $\chi$ from $140$ to $200$.
These results are reasonable as the system would get more entangled with the bath for strong coupling. For the spin-boson model, the Rabi-type system-bath coupling is able to generate a large number of excitations in the bath due to the counter-rotating terms, especially for strong coupling.
For the JC spin-boson model, since the total number of excitations is conserved, the total amount of entanglement that can be generated is largely restricted.

In addition, we observe that at $\alpha=0.01, 0.04$, the extended TEMPO results for $\beta=5$ and $\beta=\infty$ are close to each other, but at $\alpha=0.08, 0.12$, the extended TEMPO results for $\beta=5$ and $\beta=\infty$ are very different from each other. This could also be attributed to the number-conserving system-bath coupling in the JC spin-model: for weak coupling the bath can hardly get excited and the low-temperature bath behaves similar to the vacuum, but for strong coupling, the bath can get excited, and the low-temperature bath becomes different from the vacuum as it can allow more excitations. Interestingly, the behaviors of the spin-boson model are almost the opposite: at weaker couplings the results for the two temperatures are different, but get closer at strong couplings, which again shows the different physics between these two models.

% In Fig.~\ref{fig:sberrordt} and Fig.~\ref{fig:sberrorchi}, we perform convergence study of our calculations in Fig.~\ref{fig:comparison} against the two hyperparameters $\delta t$ and $\chi$, respectively. Generally, we can see that these two sets of results can easily converge against $\delta t$ and $\chi$ for small $\alpha$, and that
% the extended TEMPO calculations for the JC spin-boson model converge much faster than the TEMPO calculations for the spin-boson model as $\delta t$ becomes smaller or $\chi$ becomes larger. 
% In particular, for the JC spin-boson model, the extended TEMPO results already converge very well at $\delta t=0.025$ and $\chi=50$ for all considered values of $\alpha$. In comparison, for the spin-boson model, the TEMPO results have required $\chi=200$ and $\delta t=0.025$ to converge for $\alpha\leq 0.08$, and have not fully converged at $\alpha =0.12$ even when decreasing $\delta t$ from $0.025$ to $0.0125$ or increasing $\chi$ from $140$ to $200$.
% These results are reasonable as the system would get more entangled with the bath for strong coupling. For the spin-boson model, the Rabi-type coupling is able to generate a large number of excitations in the bath due to the counter-rotating terms, especially for strong coupling.
% For the JC spin-boson model, since the total number of excitation is conserved and we have considered a system-bath initial state in the single-excitation sector, the total amount of entanglement that can be generated is largely restricted.

\subsubsection{Imaginary-time evolution of an interacting bosonic impurity coupled to a subohmic bath}
\begin{figure*}
  \includegraphics[width=2\columnwidth]{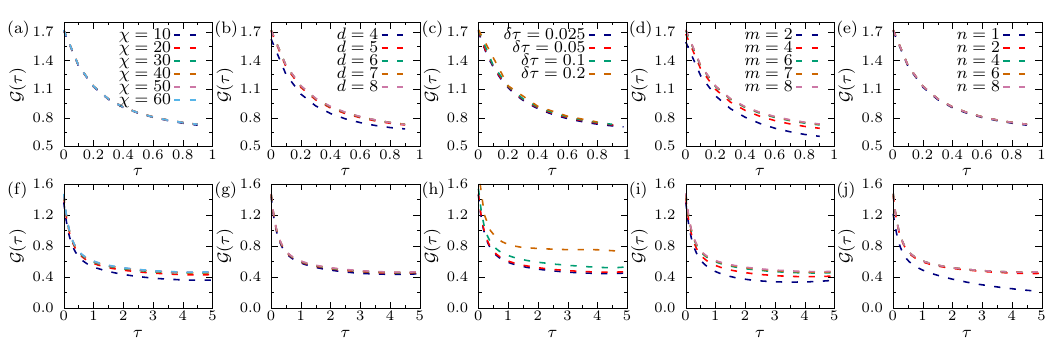}
  \caption{The Matsubara Green's function $\mathcal{G}(\tau)$ as a function of the imaginary time $\tau$ for an interacting bosonic impurity that is off-diagonally coupled to a subohmic bath, with the system Hamiltonian in Eq.(\ref{eq:bosonimpurity}). Top row: $\mathcal{G}(\tau)$ as a function of $\tau$ against the hyperparameters $\chi$ (a), $d$ (b), $\delta\tau$ (c), $m$ (d) and $n$ (e) for $\beta=1$. Bottom row: similar to the top row, but for $\beta=5$ instead. The default hyperparameters are chosen as $\chi=30$, $d=6$, $\delta\tau=0.1$ for the extended TEMPO simulations in the top row, and as $\chi=60$, $d=8$, $\delta\tau=0.05$ for the extended TEMPO simulations in the bottom row.
    }
    \label{fig:bosonic}
\end{figure*}

\begin{figure}
  \includegraphics[]{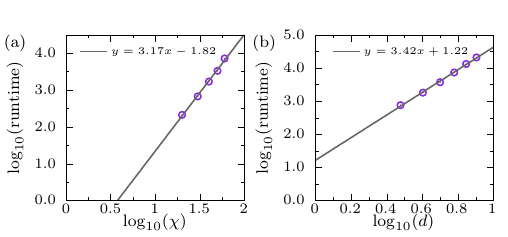}
  \caption{The runtime to generate the MPO-IF for the bosonic QIP considered in Fig.~\ref{fig:bosonic} as a function of the bond dimension $\chi$ (a) and of the local Hilbert space size $d$ (b). In (a) we have fixed $d=4$ and used $\chi=20,30,40,50,60$, in (b) we have fixed $\chi=40$ and used $d=3,4,5,6,7,8$. The purple circles are the data points and the gray solid lines are polynomial fittings. In all these calculations we have used $\beta=5$ and $\delta\tau=0.05$. These tests are obtained using a single thread on a CPU with 2.4 GHz frequency.
    }
    \label{fig:scaling}
\end{figure}

So far we have only considered the real-time evolution in both the theoretical derivations and the numerical examples. As our method takes the Feynman-Vernon IF as the only starting point, it can also be straightforwardly applied to the imaginary contour or even the L-shaped Kadanoff contour (see Refs.~\cite{ChenGuo2024b,ChenGuo2024g} for the calculations on these contours in the fermionic case using GTEMPO). In fact, the expression of the IF in Eq.(\ref{eq:IF}) is universal for all different contours. For the other contours, the only essential change to make is the boundary condition in evaluating $\gK$, for example, on the imaginary contour one has periodic boundary and no system initial state is required. 
As a final numerical example, we consider the imaginary-time evolution of an interacting bosonic impurity coupled to the subohmic bath in Eq.(\ref{eq:subohmic}) via $\Aop=\aop$, and the system Hamiltonian is chosen as
\begin{align}\label{eq:bosonimpurity}
\Himp = \nop + U\nop^2,
\end{align}
with $U$ the on-site interaction strength. In our numerical simulations of this model, we will fix $U=1$ and $\alpha=0.04$, and focus on calculating the bosonic Matsubara Green's function (which is relevant in the BDMFT iterations on the imaginary-time axis). We use the convention
\begin{align}
\mathcal{G}(\tau) = \langle \mathcal{T}_\tau \aop(\tau) \adop(0) \rangle_{\rm th} = 
\begin{cases}
\langle \aop(\tau) \adop(0) \rangle_{\rm th}, & 0 < \tau \le \beta, \\
-\langle \adop(0) \aop(\tau) \rangle_{\rm th}, & -\beta < \tau \le 0,
\end{cases}
\end{align}
where $\mathcal{T}_\tau$ is the imaginary-time-ordering operator, $\langle \cdots\rangle_{\rm th}$ denotes the thermal average, and $\aop(\tau)$ is the annihilation operator in the Heisenberg picture at imaginary time $\tau$. For $\tau>0$ we therefore evaluate $\langle \aop(\tau) \adop(0) \rangle_{\rm th}$. 

The results are plotted in Fig.~\ref{fig:bosonic}, where we have considered two temperatures: $\beta=1$ (the top row) and $\beta=5$ (the bottom row). For each temperature, we have shown the results for the five hyperparameters $\chi$, $d$, $\delta \tau$, $m$ and $n$ from left to right that may affect the numerical accuracy. Generally, we can see that the results for $\beta=1$ converge more quickly against the hyperparameters than those for $\beta=5$, except against $d$. Concretely, 
for $\beta=1$ the extended TEMPO results converge fairly well with $\chi=10$, $\delta\tau=0.1$, $m=6$ and $n=1$, while for $\beta=5$ they only converge well with $\chi=20$, $\delta\tau=0.05$, $m=6$ and $n=4$. In comparison, $d=5$ and $d=4$ are required for $\beta=1$ and $\beta=5$ respectively to obtain converged results.
These results are reasonable as the high-temperature case is generally expected to be easier to simulate than the low-temperature case on the imaginary contour, which is also supported by the fact that $n=1$ suffices for $\beta=1$ from Fig.~\ref{fig:bosonic}(e). Meanwhile, for bosonic particles one would need a larger local Hilbert space at higher temperature to accommodate for more particles. This application illustrates that our method could be straightforwardly used as an imaginary-time impurity solver in BDMFT.

It could also be insightful to look at the practical runtime scaling of our method versus the hyperparameters, for which we will use the imaginary-time evolution of the bosonic QIP considered above as the test ground. The dominant computational cost in our method to build the MPO-IF is the $m$ MPO-MPO multiplications in the XTRG algorithm, which would scale linearly with $m$ but can generally converge very quickly with $m$. For a fixed $\beta$, the number of imaginary-time steps scales linearly as $1/\delta\tau$, therefore the runtime should generally scale linearly as $1/\delta\tau$ if the other hyperparameters are the same (here we have assumed that the iterative MPO-MPO multiplication scheme requires a constant number of iterations to converge). In addition, $n$ only affect the construction of $\Heff$ as an MPO, and the cost of which is negligible for moderate values of $n$. Therefore, we will focus on the two hyperparameters $\chi$ and $d$ which has most significant effects on the performance of our method in practice. 

In Fig.~\ref{fig:scaling}(a,b), we plot the runtime scaling of our method to build the MPO-IF used in solving the bosonic QIP in Fig.~\ref{fig:bosonic} versus $\chi$ and $d$ respectively. We can see that the scaling in both cases can be well approximated by polynomial fittings (the fitting parameters are shown in the plots). The exponents of the polynomial fittings are both between $3$ and $4$, close to but not exactly their theoretical values (the theoretical scaling is $O(N\chi^4d^3)$), which are likely because the theoretical values have only considered the leading contribution, and that the considered values of $\chi$ and $d$ are not large enough to reach the asymptotic scaling.

\section{Summary}\label{sec:summary}

In summary, we have proposed an extension of the TEMPO method to the case of off-diagonal system-bath coupling where the system is coupled to the bath via a conjugate pair of non-Hermitian operators, by drawing the intimate connection between TEMPO and the process tensor framework. 
% Our method applies to any QIPs in which the system is linearly coupled to a noninteracting bosonic bath such that the analytic expression of the Feynman-Vernon IF can be derived. 
Similar to TEMPO, our method starts from a generalized analytic expression of the Feynman-Vernon influence functional in the continuous-time limit, but the generalized IF is in terms of operator paths instead of variable paths.
Our method naturally contains the previous developments of TEMPO under more restricted settings. To illustrate our method, we apply it to study a modified spin-boson model with Jaynes-Cummings-type coupling to a subohmic bath, in which we show that the physics is generally very different from that of the standard spin-boson model for the considered structural bath, even in the weak-coupling regime. 
% These results also demonstrate that the rotating-wave approximation, which is often made for the spin-boson model, could be a very bad approximation in presence of a structural bath. 
We also apply our method to study the imaginary-time evolution of an interacting bosonic impurity coupled to a subohmic bath, to illustrate the flexibility of our method to deal with different types of impurities as well as problems on different contours.
Our method provides a unified understanding of the current developments of the TEMPO method so far, and the central idea of it may also be generalized to the fermionic case. As an impurity solver for the normal phase with a scalar hybridization function, our method could serve as a building block for the bosonic dynamical mean field theory.

\textit{Note added.} The implementation of the extended TEMPO method in this work on both the imaginary-time and real-time axis, as well as the standard TEMPO, can be found at~\cite{TEMPO} (in julia language). During our preparation of this work, we notice a recent work which proposed an alternative strategy to extend TEMPO for off-diagonal system-bath coupling~\cite{Link2026}.

\textit{Data availability statement.} The data that support the findings of this article are openly available \cite{EXTENDED-TEMPO-DATA}.

\begin{acknowledgments}
This work is supported by the National Natural Science Foundation of China under Grant No. 12104328, No.12574403, No. 12174447, No. 22542010, No. 62471478 and No. 12305049. 
\end{acknowledgments}

\appendix

\section{Derivation of the Feynman-Vernon IF for off-diagonal system-bath coupling}\label{app:derivation}

Here we derive Eq.(\ref{eq:Zimp}) in the main text via the superoperator method
as described in
Refs.~\cite{ferialdi2016-exact,strathearn2020-modelling,gribben2022-exact}.
First we consider the case of a single-mode bath with the total Hamiltonian:
\begin{align}
  \Hop=\Hop_0+\Vop=\HS+\HB+\Vop,
\end{align}
where 
\begin{align}
  \HB=\omega_0\bdop\bop,\quad
  \Vop=\gamma_0(\Adop\bop+\Aop\bdop).
\end{align}
Here $\Adop,\Aop$ are system operators, $\bdop$ and $\bop$ are the bosonic creation and annihilation operators of the bath. For simplicity, in the following we assume the system-bath coupling strength (here is $\gamma_0$) to be real. 
In the interaction picture, the density matrix $\rhoop^I(t)$ satisfies the Liouville equation:
\begin{align}\label{eq:liouvillian}
  \pdv{\rhoop^I(t)}{t}=\Lop^I(t)\rhoop^I(t) = -\im[\Vop^I(t),\rhoop^I(t)],
\end{align}
where $\Vop^I(t)=e^{i\Hop_0t}\Vop e^{-i\Hop_0t}$,  $\Lop^I(t)$ is the Liouville superoperator in the interaction picture. 
The initial state $\rhoop_0$ of the system and
bath is assumed to be separable:
\begin{align}
  \rhoop_0=\rhoimp \otimes \rhobath.
\end{align}
% Defining the Liouville superoperator $\Lop^I(t)$ in the interaction picture as
% \begin{align}
%   \label{eq:liouvillian}
%   \Lop^I(t)\rhoop^I(t)=-\im[\Vop^I(t),\rhoop^I(t)],
% \end{align}
The solution $\rhoop^I(t)$ of Eq.(\ref{eq:liouvillian}) can be formally written as
\begin{align}
  \rhoop^I(t)=U^I(t,0)\rhoop_0,
\end{align}
where the evolutionary operator $U^I(t,0)$ is 
\begin{align}
  U^I(t_1,t_2)=\Top  e^{\int_{t_1}^{t_2}\Lop^I(t')\dd{t'}}.
\end{align}
Here the time-ordering operator $\Top$ organizes the superoperator with smaller time to the right. The
reduced density matrix of the system can be obtained as
\begin{align}
  \rhoop_S^I(t)=\frac{\Tr_B[\rhoop^I(t)]}{\Zbath}=U_S^I(t,0)\rhoimp,
\end{align}
where $\Zbath$ is the free bath partition function and $U_S^I$ is defined as
\begin{align}\label{eq:IF2}
  U_S^I(t,0)=\frac{\Tr_B[\Top e^{\int_0^t\Lop^I(t')\dd{t'}}\rhobath]}{\Zbath}.
\end{align}

In the following, we show that $U_S^I(t,0)\rhoimp$ defined in Eq.(\ref{eq:IF2}) is exactly the term inside the square bracket of Eq.(\ref{eq:Zimp}), i.e., 
\begin{align}\label{eq:goal}
%U_S^I(t,0)\rhoimp = \Topc \gI[\boldAdop,\boldAop]\gK[\Himp].
U_S^I(t,0)\rhoimp = \gI[\boldAdop,\boldAop]\gK[\Himp].
\end{align}
% is exactly the influence function we defefined in Eq.(\ref{eq:IF}) of the main text.
% The influence functional is defined as
% \begin{equation}
%   \gI[\aop(t),\adop(t)]=\expval{Te^{\int_0^t\Lop(t')\dd{t'}}}_B,
% \end{equation}
% where $\expval{\cdots}_B=\frac{1}{Z^{(0)}_B}\Tr_B[\cdots]$. 
In fact, $U_S^I(t,0)$ is a moment
generating functional that can be expanded as
\begin{align}
  U_S^I(t,0)=\sum_{n=0}^{\infty}\frac{1}{n!}\expval{\Top\qty(\int_0^t\Lop^I(t')\dd{t'})^n}_B,
\end{align}
which can be alternatively expressed in terms of a cumulant generating functional $\phi(t,0)=\ln U_S^I(t,0)$:
\begin{align}
\phi(t, 0)=
  \sum_{n=0}^{\infty}\frac{1}{n!}\cumulant{\Top\qty(\int_0^t\Lop^I(t')\dd{t'})^n}_B.
\end{align}
The zeroth order cumulant vanishes, the first-order cumulant is
identical to the first-order moment, and the second-order cumulant is related to the
first- and second-order moments via
\begin{align}
    &\cumulant{\Top\qty(\int_0^t{\Lop^I(t')}\dd{t'})^2}_B \nonumber \\
    =&\expval{\Top\qty(\int_0^t{\Lop^I(t')}\dd{t'})^2}_B -\expval{\Top\int_0^t{\Lop^I(t')}\dd{t'}}^2_B.
\end{align}
As $\rhobath$ is a Gaussian state,
cumulants of order higher than $2$ vanishes~\cite{ferialdi2016-exact,strathearn2020-modelling,gribben2022-exact}. Moreover, since
$\expval{\bop(t)+\bdop(t)}_B=0$ for thermal state, the first-order cumulant also vanishes. Therefore we have
\begin{align}
  \phi(t, 0)=&\frac{1}{2}\expval{\Top\qty(\int_0^t{\Lop^I(t')}\dd{t'})^2}_B \nonumber \\
    =&\frac{1}{2}\expval{\Top\int_0^t\dd{t_1}\int_0^t\dd{t_2}\Lop^I(t_1)\Lop^I(t_2)}_B.
\end{align}
Substituting the Liouville operator in Eq.(\ref{eq:liouvillian}) into the above equation, we
get
\begin{widetext}
\begin{align}
  \label{eq:liouvillian-rho}
    \frac{1}{2}\Top\Lop^I(t_1)\Lop^I(t_2)\rhoimp=-\frac{1}{2}\Top\Vop^I(t_1)\Vop^I(t_2)\rhoimp-\frac{1}{2}\rhoimp\bar{\Top}\Vop^I(t_2)\Vop^I(t_1) 
    +\frac{1}{2}\Vop^I(t_1)\rhoimp\Vop^I(t_2)+\frac{1}{2}\Vop^I(t_2)\rhoimp\Vop^I(t_1),
\end{align}
where $\Top$ is the time-ordering operator that organizes the superoperator with smaller time to the left, and the $\bar{\Top}$ is the anti-time-ordering operator that organizes the smaller time one to the right. 
The first term in Eq.(\ref{eq:liouvillian-rho}) is
\begin{align}
 \label{eq:first-term} -\frac{\gamma^2_0}{2}\int_0^t\dd{t_1}\int_0^t\dd{t_2}[\Top\AdIop(t_1)\AIop(t_2)\rhoimp]\expval{\Top\bIop(t_1)\bdIop(t_2)}_B
  -\frac{\gamma^2_0}{2}\int_0^t\dd{t_1}\int_0^t\dd{t_2}[\Top\AIop(t_1)\AdIop(t_2)\rhoimp]\expval{\Top\bdIop(t_1)\bIop(t_2)}_B.
\end{align}
The $t_1\ge t_2$ part of Eq.(\ref{eq:first-term}) is
\begin{align}
  -\frac{\gamma^2_0}{2}\int_0^t\dd{t_1}\int_0^{t_1}\dd{t_2}[\AdIop(t_1)\AIop(t_2)\rhoimp]\expval{\bIop(t_1)\bdIop(t_2)}_B
  -\frac{\gamma^2_0}{2}\int_0^t\dd{t_1}\int_0^{t_1}\dd{t_2}[\AIop(t_1)\AdIop(t_2)\rhoimp]\expval{\bdIop(t_1)\bIop(t_2)}_B,
\end{align}
swapping the variable $t_1,t_2$ in the second term above, we obtain
\begin{align}
  -\frac{\gamma^2_0}{2}\int_0^t\dd{t_1}\int_0^{t_1}\dd{t_2}[\AdIop(t_1)\AIop(t_2)\rhoimp]\expval{\bIop(t_1)\bdIop(t_2)}_B
  -\frac{\gamma^2_0}{2}\int_0^t\dd{t_2}\int_0^{t_2}\dd{t_1}[\AIop(t_2)\AdIop(t_1)\rhoimp]\expval{\bdIop(t_2)\bIop(t_1)}_B.
\end{align}
Similarly, the $t_1<t_2$ part of Eq.(\ref{eq:first-term}) can be written as
\begin{align}
  -\frac{\gamma^2_0}{2}\int_0^t\dd{t_2}\int_0^{t_2}\dd{t_1}[\AIop(t_2)\AdIop(t_1)\rhoimp]\expval{\bdIop(t_2)\bIop(t_1)}_B
  -\frac{\gamma^2_0}{2}\int_0^t\dd{t_1}\int_0^{t_1}\dd{t_2}[\AdIop(t_1)\AIop(t_2)\rhoimp]\expval{\bIop(t_1)\bdIop(t_2)}_B.  
\end{align}
Following the convention of Keldysh formalism~\cite{Keldysh1965,LifshitzPitaevskii1981,AokiWerner2014,Chen2025}, we view a time $t$ on the 
left hand side of $\rhoimp$ as a forward time and denote it by $t^+$. Then the above expression can be summed
in a compact form as (here $\rhoimp$ is dropped since it does not enter
$\phi(t, 0)$)
\begin{align}
  -\gamma_0^2\int_0^t\dd{t_1^+}\int_0^t\dd{t_2^+}\Top\AdIop(t_1^+)\AIop(t_2^+)\expval{\Top\bIop(t_1^+)\bdIop(t_2^+)}_B.
\end{align}
Similarly, we view a time $t$ on the right hand side of $\rhoimp$ as a backward time
and denote it by $t^-$, then the second term in Eq.(\ref{eq:liouvillian-rho}) gives
\begin{align}
  -\gamma_0^2\int_0^t\dd{t_1^-}\int_0^t\dd{t_2^-}\bar{\Top}\AdIop(t_1^-)\AIop(t_2^-)\expval{\bar{\Top}\bIop(t_1^-)\bdIop(t_2^-)}_B,
\end{align}
The last two terms in Eq.(\ref{eq:liouvillian-rho}) gives
\begin{align}
  \gamma_0^2\int_0^t\dd{t_1^+}\int_0^t\dd{t_2^-}\AdIop(t_1^+)\AIop(t_2^-)\expval{\bdIop(t_2^-)\bIop(t_1^+)}_B
  +\gamma_0^2\int_0^t\dd{t_1^-}\int_0^t\dd{t_2^+}\AdIop(t_1^-)\AIop(t_2^+)\expval{\bIop(t_1^-)\bdIop(t_2^+)}_B.
\end{align}

\end{widetext}
Employing the Keldysh contour formalism, we can recast the above expressions in a compact
form as
\begin{align}\label{eq:generating}
  \phi(t, 0)=-\Top_{\mathcal{C}}\int_{\contour}\dd{t_1}\int_{\contour}\dd{t_2}\AdIop(t_1)\Delta(t_1,t_2)\AIop(t_2),
\end{align}
where $\mathcal{C}$ is the Keldysh contour $0\rightarrow t\rightarrow0$. The
term $\Delta(t_1,t_2)$ is the correlation function defined as:
\begin{align}\label{eq:Delta1}
  \Delta(t_1,t_2)=\im\gamma_0^2D(t_1,t_2),
\end{align}
where $D$ is the free bath contour-ordered Green's function:
\begin{align}
  D(t_1,t_2)=-\im\expval{\Top_{\mathcal{C}}\bIop(t_1)\bdIop(t_2)}_B.
\end{align}
Therefore we arrive at the following expression for $U_S^I(t,0)$
\begin{align}
  \label{eq:Uop}
  U_S^I(t,0)=\Top_{\mathcal{C}}e^{-\int_{\contour}\dd{t_1}\int_{\contour}\dd{t_2}\AdIop(t_1)\Delta(t_1,t_2)\AIop(t_2)}.
\end{align}
To relate Eq.(\ref{eq:Uop}) to Eq.(\ref{eq:Zimp}), we take closer look at each term in the expansion of Eq.(\ref{eq:Uop}). The first-order expansion contains terms like (neglecting the double integral and the coefficient $\Delta(t_1,t_2)$):
\begin{align}
  \Top_{\mathcal{C}}\AdIop(t_1)\AIop(t_2)=\Top_{\mathcal{C}}\Adop(t_1)\Aop(t_2)e^{-\im\int_{\mathcal{C}}\dd{\tau}\Himp(\tau)}.
\end{align}
The situation are similar for higher order expansions. 
% \begin{align}
% &\AdIop(t_1)\AIop(t_2) \nonumber \\ 
%  =& e^{\im \Himp t_1}\Adop(t_1)  e^{-\im \Himp (t_1-t_2)} \Aop(t_2) e^{-\im \Himp t_2} \nonumber \\
% =& e^{\im \Himp t_1}\Adop(t_1) e^{-\im \Himp (t_1-t)}  e^{-\im \Himp (t-t_2)} \Aop(t_2) e^{-\im \Himp t_2} \nonumber \\
% =& \Topc  \Adop(t_1) \Aop(t_2) e^{-\im\int_{\contour}\dd\tau \Himp },
% \end{align}
% therefore we can rewrite Eq.(\ref{eq:generating}) as
% \begin{align}
%  \phi(t, 0)=-\Topc e^{-\im\int_{\contour}\dd\tau \Himp } \int_{\contour}\dd{t_1}\int_{\contour}\dd{t_2}\Adop(t_1)\Delta(t_1,t_2)\Aop(t_2).
% \end{align}
% where in the second equation we have extended the bare system propagators from $0\rightarrow t_2\rightarrow t_1\rightarrow 0$ to $0\rightarrow t_2\rightarrow t\rightarrow t_1 \rightarrow 0$ such that the full Keldysh contour is spanned, in the third equation we have explicitly added the contour-ordering operator such that we can move all the bare system propagators out (i.e., the factor $e^{-\im\int_{\contour}\dd\tau \Himp }$).Now we can see that by adding $\Topc$ in the front, one can move a factor $e^{-\im\int_{\contour}\dd\tau \Himp }$ out of the first-order expansion of $U_S^I(t,0)$, moreover, it is straightforward to see this holds to any order.
As a result, we can rewrite Eq.(\ref{eq:Uop}) as
\begin{align}
U_S^I(t,0)=\Top_{\mathcal{C}} e^{-\int_{\contour}\dd{t_1}\int_{\contour}\dd{t_2}\Adop(t_1)\Delta(t_1,t_2)\Aop(t_2)} e^{-\im\int_{\contour}\dd\tau \Himp } ,
\end{align}
in which all the operators are in the Schr$\ddot{\text{o}}$dinger picture. If we make the convention that $\Adop(t_1)$ and $\Aop(t_2)$ acts on history Hilbert space at time $t_1$ and $t_2$, then the contour ordering is automatically satisfied and the contour ordering operator $\Topc$ can be omitted. Therefore Eq.(\ref{eq:goal}) is proven.

If a continuous bath is considered with
\begin{align}
  \HB=\sum_k\omega_k\bdop_k\bop_k,\quad \Vop=\sum_k(V_k\Adop\bop_k+\hc),
\end{align}
we only need to replace $\Delta(t_1,t_2)$ in Eq.(\ref{eq:Delta1}) by 
\begin{align}
  \Delta(t_1,t_2)=\im\int\dd{\omega}J(\omega)D_{\omega}(t_1,t_2),
\end{align}
where $J(\omega)$ is the bath spectral function
\begin{align}
  J(\omega)=\sum_kV_k^2\delta(\omega-\omega_k).
\end{align}
For Jaynes-Cummings-type coupling, we have $\Aop=\sgm$ and $\Adop=\sgp$, and for
Rabi-type coupling, we have $\Aop=\Adop=\sgx$.

\section{Feynman-Vernon IF for multiple baths}\label{app:multiplebaths}
For simplicity, let us first consider two single-mode baths with
\begin{equation}
  \HB=\sum_{\alpha=1,2}\omega_{\alpha}\bdop_{\alpha}\bop_{\alpha},\quad \Vop=\sum_{\alpha=1,2}\gamma_{\alpha}(\Adop_{\alpha}\bop_{\alpha}+\Aop_{\alpha}\bdop_{\alpha}).
\end{equation}
Following the same procedure as described in Appendix.~\ref{app:derivation}, we have
\begin{equation}
  \phi(t,0)=\phi_1(t,0)+\phi_2(t,0),
\end{equation}
where
\begin{equation}
  \phi_{\alpha}(t,0)=-\Topc\int_{\contour}\dd{t_1}\int_{\contour}\dd{t_2}\AdIop_{\alpha}(t_1)\Delta_{\alpha}(t_1,t_2)\AIop_{\alpha}(t_2).
\end{equation}
Here $\Delta_{\alpha}$ is the hybridization function due to the $\alpha$th bath:
\begin{equation}
  \Delta_{\alpha}=\im\gamma_{\alpha}^2D_{\alpha}(t_1,t_2),
\end{equation}
with $D_{\alpha}$ the corresponding free contour-ordering Green's function. As discussed in the above section, the $\Topc$ notation can be omitted, and  the influence functional can be written as
\begin{equation}
  \mathcal{I}=e^{S_1+S_2},
\end{equation}
where
\begin{equation}
  \label{eq:S-alpha}
  S_{\alpha}=-\int_{\mathcal{C}}\dd{t_1}\int_{\mathcal{C}}\dd{t_2}\Adop_{\alpha}(t_1)\Delta_{\alpha}(t_1,t_2)\Aop_{\alpha}(t_2).
\end{equation}
Here $\Adop_k(t_1)$ and $\Aop_k(t_2)$ act on the history Hilbert space at $t_1$ and $t_2$, respectively, and $S_k$ is understood as an operator acting on the whole history Hilbert space.

At first glance, $S_1$ and $S_2$ do not commute, and we cannot write $e^{S_1+S_2}=e^{S_1}e^{S_2}$. However, since $S_k$ is in the form of double integral of $t$, the commutator $[S_1,S_2]$ will be in the form of a quadruple integral of $t$. Since the non-commutativity of $\Adop,\Aop$ only occur at equal time, the integrand in $[S_1,S_2]$ only survives on a manifold with dimension less than 4, i.e., in the $\delta t\to0$ limit $S_1,S_2$ commutes. Therefore we can separate the IF due to different bath that $e^{S_1+S_2}=e^{S_1}e^{S_2}$, and construct $e^{S_1}$ and $e^{S_2}$ separately.

In general, we have multiple baths with continuous modes that
\begin{equation}
  \HB=\sum_{\alpha,k}\omega_{\alpha k}\bdop_{\alpha k}\bop_{\alpha k},\quad,
  \Vop=\sum_{\alpha,k}\gamma_{\alpha k}(\Adop_{\alpha}\bop_{\alpha k}+\Aop_{\alpha}\bdop_{\alpha k}).
\end{equation}
Following the same procedure, we shall have
\begin{equation}
  \gI=e^{\sum_{\alpha}S_{\alpha}},
\end{equation}
where $S_{\alpha}$ has the same form as Eq.(\ref{eq:S-alpha}), the difference is that $\Delta_{\alpha}$ is replaced by
\begin{equation}
  \Delta_\alpha(t_1,t_2)=\im\sum_{k}\gamma_{\alpha k}^2D_{\alpha k}(t_1,t_2).
\end{equation}
Define the spectral function of $\alpha$th bath as
\begin{equation}
  J_\alpha(\omega)=\sum_k\gamma_{\alpha k}^2\delta(\omega-\omega_{\alpha k}),
\end{equation}
we can write
\begin{equation}
  \Delta_{\alpha}(t_1,t_2)=\im\int\dd{\omega}J_\alpha(\omega)D_{\alpha\omega}(t_1,t_2).
\end{equation}
\section{The QuAPI scheme}\label{app:quapi}
\subsection{Keldysh contour}
First let us consider the situation on the Keldysh contour. For diagonal system-bath coupling, we can discretize the Feynman-Vernon IF with $t=N\Delta t$ as
\begin{align}
  e^{-\sum_{\zeta\zeta'=\pm}\sum_{j,k}\Adop_{j,\zeta}\Delta_{j,k}^{\zeta\zeta'}\Aop_{k,\zeta'}},
\end{align}
where
\begin{align}
  \label{eq:rabi-delta}
  \Delta^{\zeta\zeta'}_{j,k}=\int_{j\Delta t}^{(j+1)\Delta t}\dd{t_1}\int_{k\Delta t}^{(k+1)\Delta t}\dd{t_2}
  \Delta^{\zeta\zeta'}(t_1,t_2).
\end{align}
Since $\Adop$ commutes with $\Aop$ in this case, there is no ambiguity at equal time step with $j=k$ and $\zeta=\zeta'$. The explicit forms of $\Delta_{j,k}^{\zeta\zeta'}$ are
\begin{widetext}
\begin{equation}
  \Delta^{++}_{j,k}=\begin{cases}
    \displaystyle 2\int\frac{J(\omega)}{\omega^2}(1-e^{-\beta\omega})^{-1}e^{-\im\omega(j-k)\Delta t}(1-\cos\omega\Delta t),&j>k,\\ 
    \displaystyle\int\dd{\omega}\frac{J(\omega)}{\omega^2}(1-e^{-\beta\omega})^{-1}[(1-\im\omega\Delta t)-e^{-\im\omega\Delta t}]
                                                                                                               +\int\dd{\omega}\frac{J(\omega)}{\omega^2}(e^{\beta\omega}-1)^{-1}[(1+\im\omega\Delta t)-e^{\im\omega\Delta t}],&j=k,\\
    \displaystyle 2\int\frac{J(\omega)}{\omega^2}(e^{\beta\omega}-1)^{-1}e^{-\im\omega(j-k)\Delta t}(1-\cos\omega\Delta t),&j<k,
  \end{cases}
\end{equation}
\begin{equation}
  \Delta^{--}_{j,k}=\begin{cases}
    \displaystyle 2\int\frac{J(\omega)}{\omega^2}(e^{\beta\omega}-1)^{-1}e^{-\im\omega(j-k)\Delta t}(1-\cos\omega\Delta t),&j>k,\\ 
    \displaystyle\int\dd{\omega}\frac{J(\omega)}{\omega^2}(e^{\beta\omega}-1)^{-1}[(1-\im\omega\Delta t)-e^{-\im\omega\Delta t}]
                                                                                                               +\int\dd{\omega}\frac{J(\omega)}{\omega^2}(1-e^{-\beta\omega})^{-1}[(1+\im\omega\Delta t)-e^{\im\omega\Delta t}],&j=k,\\
    \displaystyle 2\int\frac{J(\omega)}{\omega^2}(1-e^{-\beta\omega})^{-1}e^{-\im\omega(j-k)\Delta t}(1-\cos\omega\Delta t),&j<k,
  \end{cases}
\end{equation}
\begin{equation}
  \Delta^{+-}_{j,k}=\displaystyle 2\int\frac{J(\omega)}{\omega^2}(e^{\beta\omega}-1)^{-1}e^{-\im\omega(j-k)\Delta t}(1-\cos\omega\Delta t),
\end{equation}
\begin{equation}
  \Delta^{-+}_{j,k}=\displaystyle 2\int\frac{J(\omega)}{\omega^2}(1-e^{-\beta\omega})^{-1}e^{-\im\omega(j-k)\Delta t}(1-\cos\omega\Delta t).
\end{equation}
\end{widetext}
However, for off-diagonal system-bath coupling, we need to consider the non-commutativity of
$\Adop$ and $\Aop$. This leads to ambiguity when $j=k$ and $\zeta=\zeta'$,
and we need to consider the order of $t_1,t_2$ within the same time duration $\Delta t$.
This ambiguity is purely a discretization artifact: in the continuum limit, the equal-time manifold $\tau=\tau'$ has measure zero, and the contour-ordering operator $\Topc$ unambiguously orders operators at distinct times. It is only after discretization that the equal-time contribution acquires a finite weight and must be resolved by the prescription below.
For $\zeta=\zeta'=+$, when $t_1>t_2$, $\Adop$ is on the left of $\Aop$, and when
$t_1<t_2$, $\Adop$ is on the right of $\Aop$. Therefore we need to replace
$e^{-\Adop_{j,+}\Delta^{++}_{j,j}\Aop_{j,+}}$ in Eq.(\ref{eq:rabi-delta}) as
\begin{widetext}
  \begin{equation}
  e^{-\Adop_{j,+}\Aop_{j,+}\int_{j\Delta t}^{(j+1)\Delta t}\dd{t_1}\int_{j\Delta t}^{t_1}\dd{t_2}\Delta^{++}(t_1,t_2)
  -\Aop_{j,+}\Adop_{j,+}\int_{j\Delta t}^{(j+1)\Delta t}\dd{t_2}\int_{j\Delta t}^{t_2}\dd{t_1}\Delta^{++}(t_1,t_2)},
\end{equation}
which is
\begin{equation}
  e^{-\Adop_{j,+}\Aop_{j,+}\int\dd{\omega}\frac{J(\omega)}{\omega^2}(1-e^{-\beta\omega})^{-1}[(1-\im\omega\Delta t)-e^{-\im\omega\Delta t}]
  -\Aop_{j,+}\Adop_{j,+}\int\dd{\omega}\frac{J(\omega)}{\omega^2}(e^{-\beta\omega}-1)^{-1}[(1+\im\omega\Delta t)-e^{\im\omega\Delta t}]}.
\end{equation}
Similarly, when $\zeta=\zeta'=-$, we need to replace $e^{-\Adop_{j,-}\Delta^{--}_{j,j}\Aop_{j,-}}$ in Eq.(\ref{eq:rabi-delta}) as
\begin{equation}
  e^{-\Adop_{j,-}\Aop_{j,-}\int_{j\Delta t}^{(j+1)\Delta t}\dd{t_1}\int_{j\Delta t}^{t_1}\Delta^{--}(t_1,t_2)
  -\Aop_{j,-}\Adop_{j,-}\int_{j\Delta t}^{(j+1)\Delta t}\dd{t_2}\int_{j\Delta t}^{t_2}\Delta^{--}(t_1,t_2)},
\end{equation}
which is
\begin{equation}
  e^{-\Adop_{j,-}\Aop_{j,-}\int\dd{\omega}\frac{J(\omega)}{\omega^2}(e^{\beta\omega}-1)^{-1}[(1-\im\omega\Delta t)-e^{-\im\omega\Delta t}]
  -\Aop_{j,-}\Adop_{j,-}\int\dd{\omega}\frac{J(\omega)}{\omega^2}(1-e^{-\beta\omega})^{-1}[(1+\im\omega\Delta t)-e^{\im\omega\Delta t}]}.
\end{equation}
\end{widetext}

\subsection{Imaginary contour}
Now we consider the imaginary contour with $\tau\in[0, \beta]$, the IF is written as
\begin{equation}
  \gI[\boldAdop,\boldAop]=e^{-\int_0^\beta\dd{\tau_1}\int_0^{\beta}\dd{\tau_2}\Adop(\tau_1)\Delta(\tau_1,\tau_2)\Aop(\tau_2)},
\end{equation}
where
\begin{equation}
  \Delta(\tau_1,\tau_2)=\int\dd{\omega}J(\omega)D_{\omega}(\tau_1,\tau_2).
\end{equation}
Here $D_{\omega}$ is the free bath Matsubara Green's function that
\begin{equation}
  D_{\omega}(\tau_1,\tau_2)=\begin{cases}
    -(1-e^{-\beta\omega})^{-1}e^{-\omega(\tau_1-\tau_2)}, & \tau_1\ge\tau_2,\\
    -(e^{\beta\omega}-1)^{-1}e^{-\omega(\tau_1-\tau_2)}, & \tau_1<\tau_2.
  \end{cases}
\end{equation}
For diagonal system-bath coupling, this IF can be discretized with $\beta=N\delta\tau$ as
\begin{equation}
  \label{eq:imag-delta}
  e^{-\sum_{j,k}\Adop_j\Delta_{j,k}\Aop_k},
\end{equation}
where
\begin{equation}
  \Delta_{j,k}=\int_{j\Delta\tau}^{(j+1)\Delta\tau}\dd{\tau_1}\int_{k\Delta\tau}^{(k+1)\Delta\tau}\Delta(\tau_1,\tau_2).
\end{equation}
The explicit forms of $\Delta_{j,k}$ are
\begin{widetext}
\begin{equation}
  \Delta_{j,k}=\begin{cases}
    \displaystyle2\int\dd{\omega}\frac{J(\omega)}{\omega^2}(1-e^{-\beta\omega})^{-1}e^{-\omega(j-k)\Delta\tau}(1-\cosh\omega\Delta\tau),& j>k,\\
    \displaystyle\int\dd{\omega}\frac{J(\omega)}{\omega^2}(1-e^{-\beta\omega})^{-1}[e^{-\omega\Delta\tau}-(1-\omega\Delta\tau)]
    +\int\dd{\omega}\frac{J(\omega)}{\omega^2}(e^{\beta\omega}-1)^{-1}[e^{\omega\Delta\tau}-(1+\omega\Delta\tau)],& j=k,\\
    \displaystyle2\int\dd{\omega}\frac{J(\omega)}{\omega^2}(e^{\beta\omega}-1)^{-1}e^{-\omega(j-k)\Delta\tau}(1-\cosh\omega\Delta \tau),&j<k.\\
  \end{cases}
\end{equation}
For off-diagonal system-bath coupling, similar to the Keldysh case, the equal-time ambiguity is a discretization artifact. We need to consider the order of $\tau_1,\tau_2$ within the same time duration $\Delta\tau$, for which we need to replace $e^{-\Adop_j\Delta_{j,j}\Aop_j}$ in Eq.(\ref{eq:imag-delta}) as
\begin{equation}
  e^{-\Adop_j\Aop_j\int_{j\Delta\tau}^{(j+1)\Delta\tau}\dd{\tau_1}\int_{j\Delta\tau}^{\tau_1}\dd{\tau_2}\Delta(\tau_1,\tau_2)-
  \Aop_j\Adop_j\int_{j\Delta\tau}^{(j+1)\Delta\tau}\dd{\tau_2}\int_{j\Delta\tau}^{\tau_2}\dd{\tau_1}\Delta(\tau_1,\tau_2)},
\end{equation}
which is
\begin{equation}
  e^{-\Adop_j\Aop_j\int\dd{\omega}\frac{J(\omega)}{\omega^2}(1-e^{-\beta\omega})^{-1}[e^{-\omega\Delta\tau}-(1-\omega\Delta\tau)]
  -\Aop_j\Adop_j\int\dd{\omega}\frac{J(\omega)}{\omega^2}(e^{\beta\omega}-1)^{-1}[e^{\omega\Delta\tau}-(1+\omega\Delta\tau)]}.
\end{equation}
\end{widetext}

\section{More details on the XTRG algorithm to build the MPO-IF}\label{app:xtrg}
The first step to build the MPO representation of $e^{-\delta\Heff}$ is to build $\Heff$ as an MPO. If done exactly, one would result in an MPO with bond dimension $\chi=2N$ in general, which is very expensive for large $N$. To obtain a compressed MPO, an effective strategy is to approximate the hybridization function $\Delta^{\zeta\zeta'}_{j-i} = \Delta^{\zeta\zeta'}_{i,j}$ as the sum of exponentially decaying functions for each $\zeta,\zeta'$. Taking a particular $\zeta,\zeta'$, we assume that
\begin{align}
\Delta^{\zeta\zeta'}_{x} = \sum_{l=1}^n \alpha_l\lambda_l^{|x|},
\end{align}
where $\alpha_l$, $\lambda_l$, $n$ are parameters that can be determined by the Prony algorithm for example~\cite{marple2019digital} (in our numerical examples we have used a fixed $n$ to control the accuracy of this approximation). The whole time interval (either real or imaginary) is used for the Prony algorithm. We note that the $\alpha_l$ and $\lambda_l$ generated in Prony algorithm will be complex in general, even for imaginary-time evolution where the hybridization function is real. Nevertheless, the whole algorithm works equally for complex numbers.
After that, one could directly write down the site tensor of the MPO representation of the part of $\Heff$ with a particular $\zeta,\zeta'$ as (a single site tensor is enough as each $\Delta^{\zeta\zeta'}_{i,j}$ is time-translationally invariant)
\begin{align}\label{eq:ttiif}
\left[\begin{array}{ccccccccc} 1 & \alpha_1 \Adop & \cdots & \alpha_n \Adop & \bar{\alpha}_1 \Aop & \cdots & \bar{\alpha}_n \Aop & \Delta_0 \Adop\Aop  \\
                 0 & \lambda_1 & \cdots & 0 & 0 & \cdots & 0 & \lambda_1 \Aop \\  
                 \vdots & \vdots & \cdots & \vdots & \vdots & \cdots & \vdots & \vdots \\  
                 0 & 0 & \cdots & \lambda_n & 0 & \cdots & 0 & \lambda_n \Aop \\  
                 0 & 0 & \cdots & 0 & \bar{\lambda}_1 & \cdots & 0 & \bar{\lambda}_1 \Adop \\
                 \vdots & \vdots & \cdots & \vdots & \vdots & \cdots & \vdots & \vdots \\   
                 0 & 0 & \cdots & 0 & 0 & \cdots & \bar{\lambda}_n & \bar{\lambda}_n \Adop \\  
                 0 & 0 & \cdots & 0 & 0 & \cdots & 0 & 1 \\  
\end{array}\right],
\end{align}
where $\alpha_l$ and $\lambda_l$ correspond to $\Delta^{\zeta\zeta'}_{i,j}$ for $i<j$, $\bar{\alpha}_l$ and $\bar{\lambda}_l$ correspond to $\Delta^{\zeta\zeta'}_{i,j}$ for $i>j$ (here $\bar{\alpha}_l$ and $\bar{\lambda}_l$ do not mean the complex conjugates of $\alpha_l$ and $\lambda_l$), and we have neglected the time step indices of the operators due to the time-translational invariance. With Eq.(\ref{eq:ttiif}) for each part of $\Heff$, one can directly write down the site tensor for the corresponding part of $e^{-\delta \Heff}$ with a particular $\zeta,\zeta'$ using the $\WI$ or $\WII$ method, which will be an MPO with bond dimension $2n+1$~\cite{ZaletelPollmann2015}. Moreover, it has been shown that $n$ generally grows very slowly for general continuous bath spectral function~\cite{VilkoviskiyAbanin2024,HuangLin2026}, and in this work we found that $n=20$ for each combination of $\zeta$ and $\zeta'$ is enough for all the cases studied. Another hyperparameter used in the XTRG algorithm is the step size $\delta =1/2^m$, determined by the integer $m$. Since XTRG converges exponentially fast with $m$, and that the effective inverse temperature of the target thermal state $e^{-\delta\Heff}$ is only $1$, we found that all our numerical simulations have well converged with $m=7$.

% \section{The zipup algorithm for MPO-MPO multiplication}\label{app:zipup}

% \bibliographystyle{apsrev4-2}
\bibliography{refs}

\end{document}